\documentclass[
  cha,
  aip,
  reprint,
  showpacs
]{revtex4-1}

\usepackage[T1]{fontenc}
\usepackage[utf8]{inputenc}
\usepackage{amsmath}
\usepackage{amssymb}
\usepackage{graphicx}
\usepackage{grffile}
\usepackage{refstyle}
\usepackage[normalem]{ulem}
\newcommand{\ag}{\textcolor{black}}

\usepackage[
  citecolor=blue,
  colorlinks,
  linkcolor=blue,
  urlcolor=blue,
]{hyperref}

\begin{document}
\title{Occasional coupling enhances amplitude death in delay-coupled oscillators}
\author{Anupam Ghosh}
\email{anupamghosh0019@gmail.com (A. G.)}
\affiliation{Department of Aerospace Engineering, Indian Institute of Technology Madras, Chennai, Tamil Nadu 600036, India}
\author{Sirshendu Mondal}
\affiliation{Department of Mechanical Engineering, National Institute of Technology Durgapur, Durgapur, West Bengal 713209, India}
\author{R. I. Sujith}
\affiliation{Department of Aerospace Engineering, Indian Institute of Technology Madras, Chennai, Tamil Nadu 600036, India}
\begin{abstract}
This paper aims to study amplitude death in time delay coupled oscillators using the occasional coupling scheme that implies the intermittent interaction among the oscillators. An enhancement of amplitude death regions (i.e., an increment of the width of the amplitude death regions along the control parameter axis) can be possible using the occasional coupling in a pair of delay-coupled oscillators. Our study starts with coupled limit cycle oscillators (Stuart-Landau) and coupled chaotic oscillators (R\"ossler). We further examine coupled horizontal Rijke tubes, a prototypical model of thermoacoustic systems. Oscillatory states are highly detrimental to thermoacoustic systems such as combustors. Consequently, a state of amplitude death is always preferred. We employ the on-off coupling (i.e., a square wave function), as an occasional coupling scheme, to these coupled oscillators. On monotonically varying the coupling strength (as a control parameter), we observe an enhancement of amplitude death regions using the occasional coupling scheme compared to the continuous coupling scheme. In order to study the contribution of the occasional coupling scheme, we perform a detailed linear stability analysis and analytically explain this enhancement of the amplitude death region for coupled limit cycle oscillators. We also adopt the frequency ratio of the oscillators and the time delay between the oscillators as the control parameters. Intriguingly, we obtain a similar enhancement of the amplitude death regions using frequency ratio and time delay as the control parameters in the presence of the occasional coupling. Finally, we use a half-wave rectified sinusoidal wave function (motivated by practical reality) to introduce the occasional coupling in time-delay coupled oscillators and get similar results.
\end{abstract}
\maketitle
\begin{quotation}
The occasional coupling scheme (OCS) is a familiar topic in the purview of synchronization. Recently, the effect of OCS has been studied in the context of amplitude death in diffusively coupled oscillators. We extend the investigation and employ the OCS in time delay coupled oscillators. Time delay coupled oscillator models are beneficial in studying various real-life events, and consequently, such models are used in different disciplines like physics, biology, and engineering. Towards this, we initially choose two different examples of low-dimensional oscillators: Stuart-Landau (a limit cycle oscillator) and R\"ossler (a chaotic oscillator). We further examine thermoacoustic systems (coupled horizontal Rijke tubes) wherein amplitude death is a state of preference. In the present study, we employ the on-off coupling (that introduces the occasional coupling through a square wave function) to these three models of coupled oscillators and observe that the amplitude death regions enhance along the control parameter axes in the presence of the OCS. The reason of this enhancement is studied analytically using a local stability analysis for coupled limit cycle oscillators. We choose the coupling strength parameter, frequency ratio, and time delay as the control parameters. We further employ a different mathematical form (a half-wave rectified sinusoidal wave function) of the OCS and reach a similar conclusion. 
\end{quotation}

\section{Introduction}
\label{sec:intro}
The coupled oscillator model is widely used to study the collective dynamics of various phenomena in different fields such as physics, mathematics, engineering, and biology~\cite{winfree01,strogatz07}. Various nonlinear phenomena, viz., synchronization, chimera state, pattern formation, swarming, and amplitude death, have been studied extensively using this model~\cite{lakshmanan03,balanov08}. Here, we use this model to study one such nonlinear phenomenon: amplitude death (AD)~\cite{saxena12}.
AD, a homogeneous steady state, implies the complete suppression of oscillations of the coupled oscillators due to the coupling between them~\cite{prasad05,saxena12}. In other words, all the interacting oscillators reach the same stable fixed point during AD. The steady state exists in the uncoupled oscillators as an unstable state, and quenching is detected as the steady state becomes stable because of the coupling among the oscillators~\cite{zou21}. Rayleigh first reported the evidence of AD in a physical system where he had placed two organ pipes side by side~\cite{rayleigh96,strutt11}. Subsequent studies report that AD has been achieved in various mathematical models~\cite{saxena12,zou21} and experimental setups~\cite{vidal81,zeyer01,manoj18}. In addition, AD in coupled oscillators can be induced through numerous methods, and some of them are the following: parameter mismatch~\cite{saxena12}, time delay coupling~\cite{reddy98,saxena10}, dynamic coupling~\cite{konishi03}, conjugate coupling~\cite{karnatak07},
mean-field diffusion~\cite{sharma12}, and nonlinear coupling~\cite{prasad10}. 
Although AD has been detected in nature and studied extensively, the occurrence of AD has many practical applications. AD finds its application in many physical systems, e.g., controlling vibration~\cite{song99} in mechanical engineering, suppressing thermoacoustic~\cite{juniper18,sujith21} and aeroelastic instability~\cite{raaj21,raj21} in aerospace engineering. In such systems, oscillations are undesirable, and AD is utilized to suppress the unwanted oscillations~\cite{thomas18,thomas18_2,raaj19,raj21}. Besides, AD could be a beneficial strategy to prevent the widespread of harmful neural activities, which may further lead to various psychological diseases like epilepsy, schizophrenia, and Parkinson’s disease~\cite{tang18}. Therefore, an enhancement of the AD region is the top priority in the aforementioned systems. Recently, Sun et al~\cite{sun18} have employed the occasional coupling scheme~\cite{ghosh20} (the on-off coupling~\cite{cqh09}, to be specific) to the interacting oscillators with parameter mismatch and ascertained that the AD regions enhance along the coupling strength parameter axis using the OCS than that using the continuous coupling scheme (CCS).
OCS involves intermittent interactions of the coupled oscillators and is advantageous over CCS in attaining synchrony at the larger values of coupling strength~\cite{cqh09,ghosh18,ghosh20}. This occasional interaction between the oscillators depends explicitly either on the evolution time or on the phase space coordinates~\cite{ghosh20}. OCS was first introduced in the context of chaotic synchronization in $1993$~\cite{gupte93}. Subsequent studies report different examples of occasional coupling schemes leading to chaotic synchronization~\cite{ghosh20}. \ag{This occasional interaction is either deterministic or stochastic. For the stochastic on-off coupling scheme~\cite{jeter15}, the switching on (or off) of the coupling term is random. On the other hand, this interaction is deterministic in the on-off coupling scheme~\cite{cqh09}. The notion of occasional coupling has been extended in a complex network with delay coupling to study synchronization~\cite{sun16}. A non-synchronizable network can be made synchronizable using the OCS~\cite{sch16}. In a complex network of multiple layers, synchronization has been studied using time-varying inter-layer links~\cite{eser21}. The OCS has also been used to overcome measure desynchronization observed in coupled Hamiltonian systems~\cite{ghosh18_2}. Synchronization in circadian oscillators in single cells of fungal systems has been studied using the stochastic intermittent coupling~\cite{deng16}. It has been reported that periodically time-varying switching of coupling among the neuron oscillators can enhance synchronization~\cite{parastesh19}. The broken symmetry in coupled Josephson junctions can be restored using the OCS~\cite{tian20}. However, we use the on-off coupling scheme in this paper. By construction, using this scheme, the coupling among the interacting oscillators activates (or deactivates) periodically. We discuss this scheme elaborately in Sec.~\ref{sec:model}.}
Although OCS is mostly scrutinized to study synchronization in coupled oscillators, Sun et al~\cite{sun18} studied the effect of OCS in the context of amplitude death in diffusively coupled oscillators. The AD region is shown to extend along the coupling strength parameter axis in the presence of the OCS~\cite{sun18}. Deriving motivation from this study, we examine the enhancement of AD regions with OCS, however, in the presence of time delay in the coupling.  
Time delay in coupling is quite natural in practical systems. This time delay arises because of the finite speed of information transfer in physical systems~\cite{sipahi11}. In the literature, the time-delay coupled oscillator models have been used to study various experimental observations, viz., candle oscillators~\cite{manoj18}, electronic circuits~\cite{reddy00}, and thermo-optical oscillators~\cite{herrero00}. Reddy \textit{et al.}~\cite{reddy98} first reported the observation of AD in time-delay coupled oscillators studying coupled Stuart-Landau (SL) oscillators. All these studies, however, have used the CCS. To the best of our knowledge, the effect of OCS has not been studied for delay-coupled oscillators. Therefore, in this paper, \emph{we investigate the effect of OCS on the phenomenon of AD in delay-coupled oscillators}.
In what follows, we employ the OCS to three pairs of delay coupled oscillators: SL, R\"ossler~\cite{roessler76}, and the horizontal Rijke tube~\cite{gopalakrishnan15,sujith21} (a prototypical model of a thermoacoustic system). Initially, the coupling strength parameter is chosen as the required control parameter. Then, deriving motivation from the physical systems, we choose frequency detuning and time delay as the control parameters. Such parameters are important in practical systems and, therefore, play crucial roles in the present study. We show that the AD region is enhanced with the employment of OCS. 
This paper is structured as follows: first, we discuss the general notion of occasional coupling in the context of two interacting oscillators (in Sec. \ref{sec:model}). Subsequently, we choose two examples of low-dimensional oscillators: coupled SL oscillators and coupled R\"ossler oscillators and study the effect of OCS on AD (in Secs.~\ref{sec:sl} and \ref{sec:ross}). Finally, we extend our investigation to a mathematical model of coupled horizontal Rijke tubes (in Sec.~\ref{sec:rijke}). In Sec.~\ref{sec:other}, the results, using the half-wave rectified sinusoidal wave as a coupling function, are presented. Finally, the major conclusions of this study are summarized in Sec.~\ref{sec:conclusion}. 
\section{Results}
\label{sec:result}
\subsection{A general model}
\label{sec:model}
In order to study AD, we use two methods --- parameter mismatch and time delay coupling --- simultaneously to couple the oscillators. In most practical situations, we do not have any liberty to use the methods separately and therefore, we need to incorporate them simultaneously~\cite{thomas18,dange19}. A slight mismatch in any mechanical parameter between two physical oscillators is inevitable. Furthermore, a finite value of time is required for the propagation of information from one oscillator to another, giving rise to a time delay in coupling. These two attributes might be unavoidable in practical oscillators and therefore, they are simultaneously employed. Thus, the general form of equations of motion of two coupled oscillators using diffusive and time delay couplings are as follow: 
\begin{subequations}
	\label{eq:1}
	\begin{eqnarray}
	\dot{\mathbf{x}}_1 &=& \mathbf{F(x}_1, \mu_1) + \alpha_d \cdot (\mathbf{x}_2 - \mathbf{x}_1) + \alpha_{\tau} \cdot (\mathbf{x}_{2 \tau} - \mathbf{x}_1),\label{eq:1a}\\
	\dot{\mathbf{x}}_2 & =& \mathbf{F(x}_2, \mu_2) + \alpha_d \cdot (\mathbf{x}_1 - \mathbf{x}_2) + \alpha_{\tau} \cdot (\mathbf{x}_{1 \tau} - \mathbf{x}_2).\label{eq:1b}
	\end{eqnarray}
\end{subequations}
The vectors $\mathbf{x}_1(t) \in \mathbb{R}^m$ (where $m$ is an integer with $m >1$) and $\mathbf{x}_2(t) \in \mathbb{R}^m$ are the $m$-dimensional phase space coordinates of the first and second oscillators, respectively. The dots represent the time derivative and $\mathbf{F}(\cdot)$ is the functional form of the oscillator. $\mu_1$ and $\mu_2$ are, respectively, parameters of the interacting oscillators. $\alpha_d$ and $\alpha_{\tau}$ are respectively the diffusive coupling strength and time-delay coupling strength. Finally, $\mathbf{x}_{1 \tau} := \mathbf{x}_{1} (t - \tau)$ and $\mathbf{x}_{2 \tau} := \mathbf{x}_{2} (t - \tau)$, and the scalar $\tau$ introduces time delay in the coupling terms. Thus, the second and third terms of either equation (Eq.~\ref{eq:1a} or \ref{eq:1b}) are respectively the diffusive and time-delay coupling terms. 
\begin{figure}[htbp!]
	\hspace*{5 mm}
	\includegraphics[width=50cm,height=4.5cm, keepaspectratio]{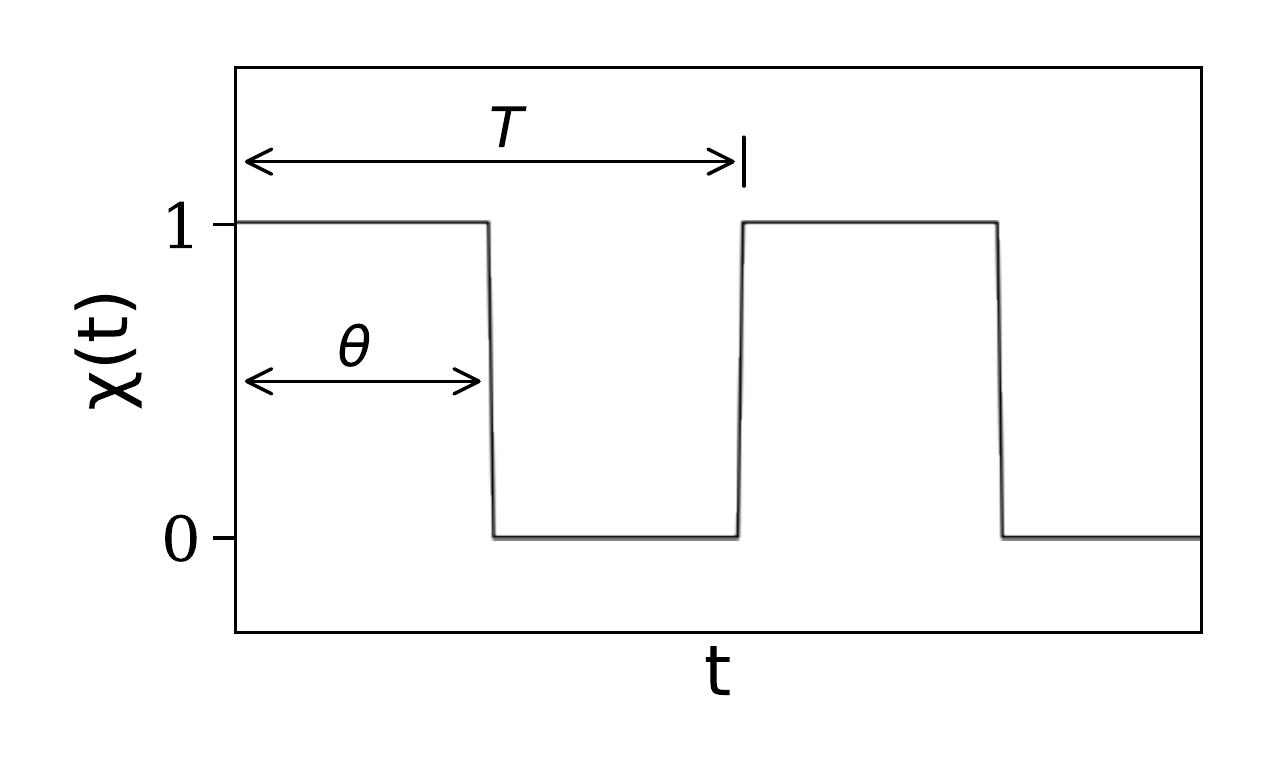}
	\caption{The schematic diagram depicts the variation of the occasional coupling function $\chi(t)$ with the increment of time $t$ using the on-off coupling scheme (Eq.~\ref{eq:ocs}).} 
	\label{fig:ocs_schematic}
\end{figure}
The on-off coupling scheme~\cite{cqh09}, a type of OCS, implies that the interaction between the oscillators $\mathbf{x}_1(t)$ and $\mathbf{x}_2(t)$ is controlled using a square wave whose amplitude switches between $0$ and $1$. The coupling between $\mathbf{x}_1(t)$ and $\mathbf{x}_2(t)$ is activated when the amplitude of the square wave is one, and both the oscillators evolve independently for zero amplitude of the square wave. Mathematically, we can define the function as follows: 
\begin{equation}
\label{eq:ocs}
\chi (t) := 
\begin{cases}
1,  \, nT < t \leq (n+\theta)T,\\
0,  \, (n + \theta)T < t \leq (n+1)T,
\end{cases}
\end{equation}
where $n$ is an integer. $T$ and $\theta \in (0, 1)$ are the on-off period and the on-off rate, respectively. Intuitively, $\theta$ measures the fraction of the time period $T$ over which the coupling term is activated. In other words, $T$ and $\theta$, respectively, imply the time period and duty cycle of the square wave. The variation of $\chi(t)$ using the on-off coupling (Eq.~\ref{eq:ocs}) is depicted schematically in Fig.~\ref{fig:ocs_schematic}. In order to employ the on-off coupling, we need to choose the parameters $T$ and $\theta$ appropriately. In the literature, the average inter-peak separation of the isolated oscillator is recommended as the typical order of $T$~\cite{chen10}. More explicitly, we plot one of the phase space coordinates as a function of time and point out the time between two consecutive local maxima (or minima), and calculate the average inter-peak interval. However, on the other hand, no such guideline is available for $\theta$ in the literature. Note that $\theta = 1$ indicates that both the oscillators are always coupled, i.e., the CCS is activated between $\mathbf{x}_1(t)$ and $\mathbf{x}_2(t)$ --- whereas $\theta = 0$ represents the uncoupled state, i.e., $\mathbf{x}_1(t)$ and $\mathbf{x}_2(t)$ are mutually independent. Thus, after incorporating OCS, Eq.~\ref{eq:1} reduces to 
\begin{subequations}
	\label{eq:1_ocs}
	\begin{eqnarray}
	\dot{\mathbf{x}}_1 =&& \mathbf{F(x}_1, \mu_1) + \alpha_d \cdot  \chi (t) \cdot (\mathbf{x}_2 - \mathbf{x}_1) \nonumber \\ &&+ \alpha_{\tau} \cdot \chi (t) \cdot (\mathbf{x}_{2 \tau} - \mathbf{x}_1),\label{eq:ocsa}\\
	\dot{\mathbf{x}}_2 =&& \mathbf{F(x}_2, \mu_2) + \alpha_d  \cdot \chi (t) \cdot (\mathbf{x}_1 - \mathbf{x}_2) \nonumber \\ &&+ \alpha_{\tau} \cdot \chi (t) \cdot (\mathbf{x}_{1 \tau} - \mathbf{x}_2).\label{eq:ocsb}
	\end{eqnarray}
\end{subequations}
Having described the general model, we turn our attention to the different examples of coupled oscillators and study the effect of OCS on AD. Note that the fourth-order Runge--Kutta method, with the smallest time step of $0.01$ and the maximum evolution time of $2000$, is used in this paper to integrate of differential equations. The time-delayed terms are considered as inactive (i.e., $\alpha_{\tau} = 0$ in Eqs.~\ref{eq:1} and \ref{eq:1_ocs}) up to $t = \tau$.
\subsection{Coupled SL oscillators}
\label{sec:sl}
We start our discussion with an example of coupled SL oscillators. SL oscillator, a type of limit cycle oscillator, is a two dimensional, autonomous dynamical system. The corresponding equations of motion are as follow~\cite{reddy98,reddy99,lakshmanan11}: 
\begin{subequations}
	\label{eq:sl}
	\begin{eqnarray}
	\dot{Z}_1 =&& \left( 1 + i \omega_1 - |Z_1|^2 \right) Z_1 + \alpha_d \cdot \chi (t) \cdot (Z_2 - Z_1) \nonumber \\ &&+ \alpha_{\tau} \cdot \chi (t) \cdot (Z_{2 \tau} - Z_1),\label{eq:sla}\\
	\dot{Z}_2 =&&  \left( 1 + i \omega_2 - |Z_2|^2 \right) Z_2 + \alpha_d \cdot \chi (t) \cdot (Z_1 - Z_2) \nonumber \\ &&+ \alpha_{\tau} \cdot \chi (t) \cdot (Z_{1 \tau} - Z_2),\label{eq:slb}
	\end{eqnarray}
\end{subequations}
where $Z_j = (x_j + i y_j), i = \sqrt{-1},$ and $j = 1,2$. Thus, for the example in hand, $\mathbf{x}_1 = (x_1, y_1)$ and $\mathbf{x}_2 = (x_2, y_2)$. The parameters $\omega_j$ are the natural frequencies of the respective SL oscillators. In the case of coupled SL oscillators, for simplicity, we choose $\alpha_d = \alpha_{\tau} = \alpha$ (say). 
\begin{figure}[htbp!]
	\hspace*{-0.36 cm}
	\includegraphics[width=70cm,height=9.3 cm, keepaspectratio]{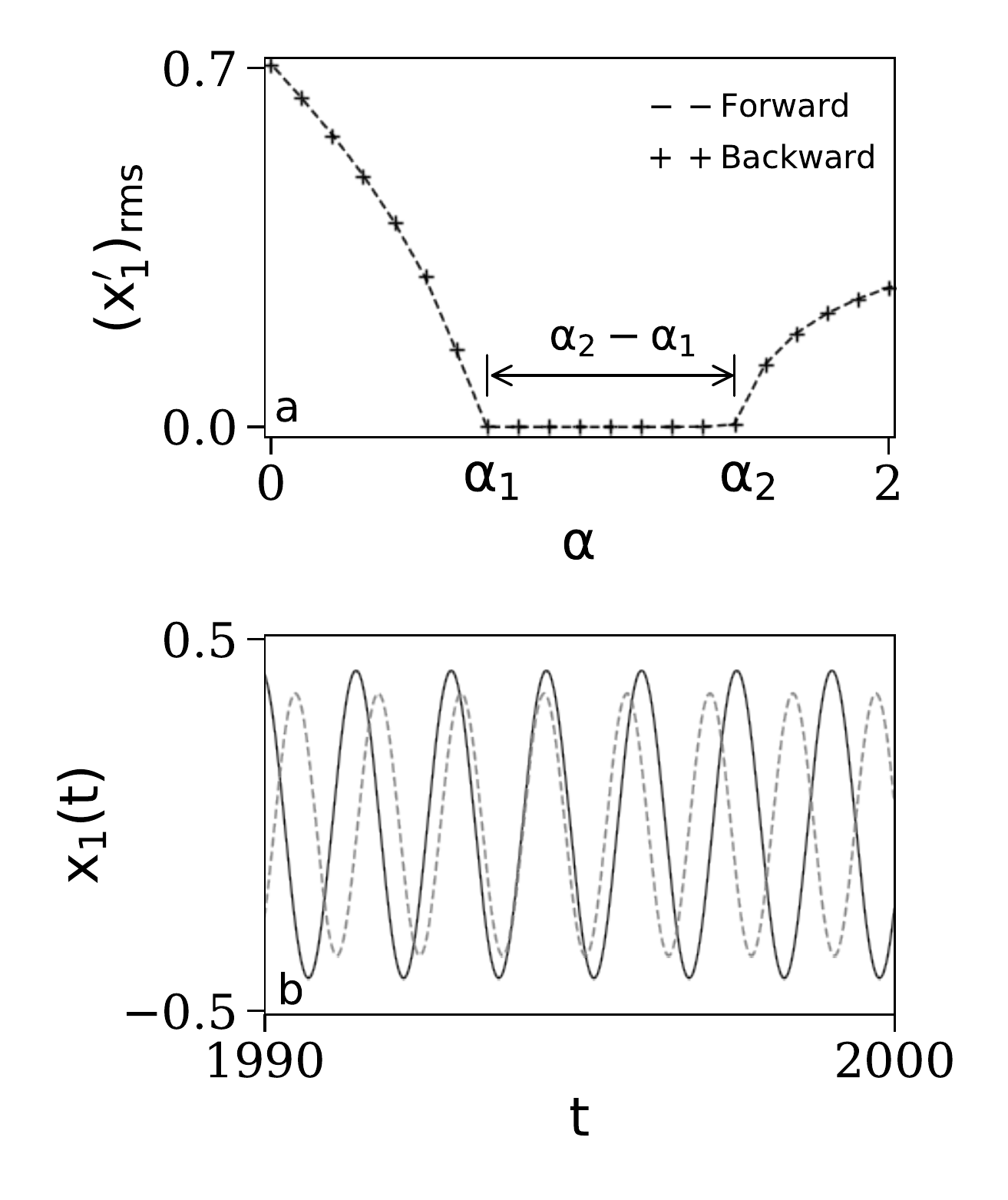}
	\caption{Transitions between LCO and AD are detected in coupled SL oscillators as we increase $\alpha$ monotonically. The time delay ($\tau$) and natural frequencies ($\omega_1$ and $\omega_2$) remain fixed in both subplots at $\tau = 0.1$, $\omega_1 = 4$, and $\omega_2 = 8$. (a) Bifurcation diagram of the coupled SL oscillators is plotted using the variable $x_1(t)$. (b) The variable $x_1(t)$ is plotted for two different values of $\alpha$: $0.5$ (black line plot) and $1.9$ (gray dashed-line plot).}
	\label{fig:sl_dyn}
\end{figure}
In order to confirm the AD region numerically, we run our code long enough, skip the transient, and take the final $10\%$ data. When the condition $|x_1(t)| < \varepsilon$ satisfies, we calculate the distance $l_{\tau_0} = |{x}_1(t)-{x}_1(t-\tau_0)|$, where $\varepsilon (= 10^{-6})$ is an arbitrary small number and $\tau_0 = 0.2$. Thus, we get an array consists of different values of $l_{\tau_0}$ with total number of elements of $N_{\tau_0}$. Finally, we take the average distance 
\begin{equation}
	\label{eq:l}
	l = (1/N_{\tau_0}) \sum_{j = 1}^{N_{\tau_0}} (l_{\tau_0})_j.
\end{equation}
If $l \leq \varepsilon$, we consider it as AD state; otherwise, it is considered as an oscillatory state. Note that a similar value of $l$ is possible to calculate using either of the other three phase space coordinates (i.e., $x_2, \, y_1,$ and $y_2$) of Eq.~\ref{eq:sl}. In all cases, we arrive at the same conclusion. Also, an AD state is independent of $\tau_0$; in principle, any value of $\tau_0$ should work. The only caution we have to consider is for the periodic motion, and $\tau_0$ must differ from the time period of oscillations. However, we use the discussed technique to ascertain the AD state numerically throughout the paper.
\begin{figure}[htbp!]
	\hspace*{-5 mm}
	\includegraphics[width=40cm, height=9cm, keepaspectratio]{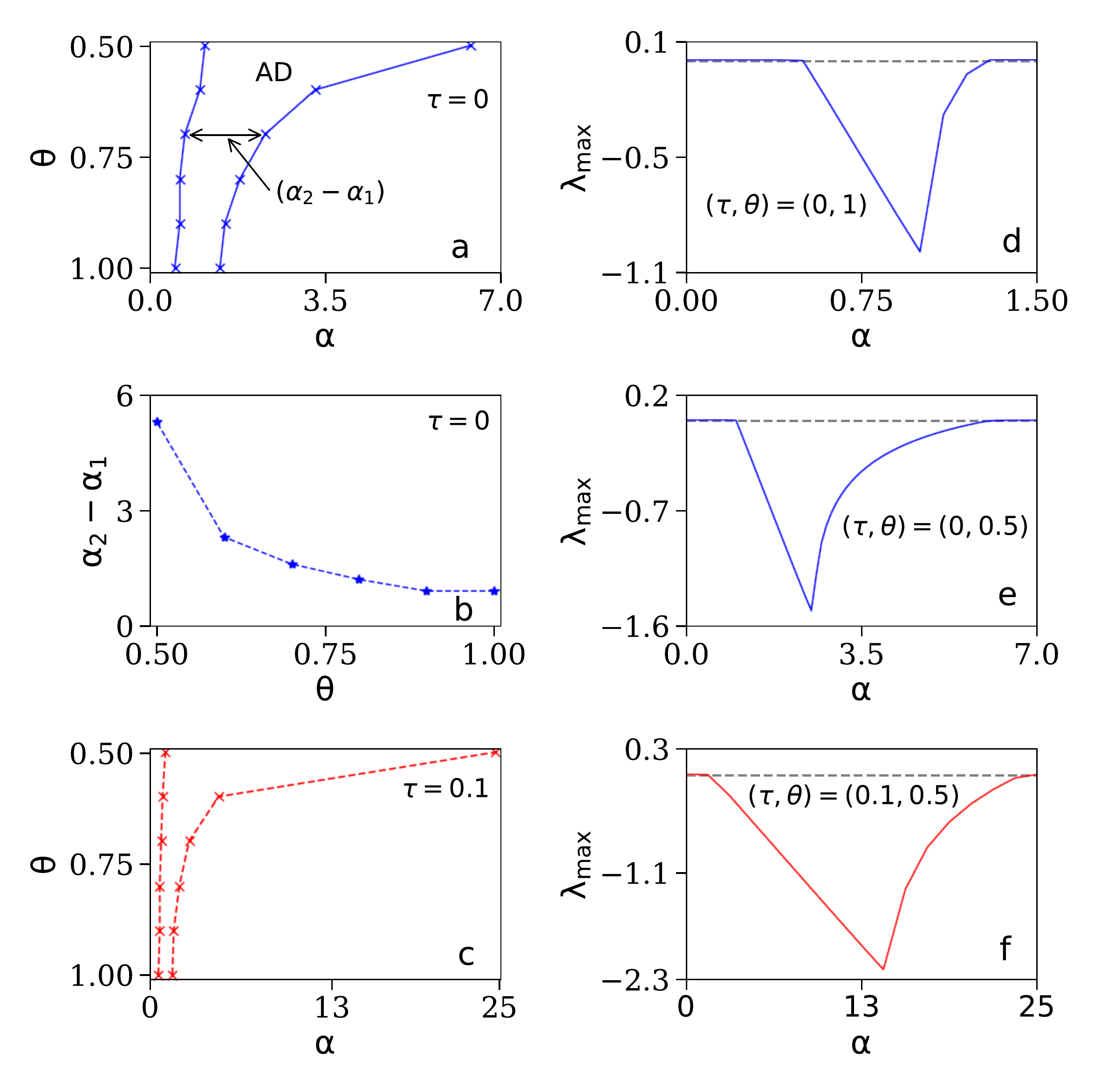}
	\caption{(\textit{Color online}) \ag{The on-off coupling is employed in coupled SL oscillators (Eq.~\ref{eq:sl}). Using two different measures: $l$ (first column) and $\lambda_{\rm max}$ (second column), we have studied the AD in the presence and absence of time delay. The width of the AD region enhances more significantly for $\tau = 0.1$ than $\tau = 0.0$ at smaller values of $\theta$. For all subplots, the on-off period $T = 1$. The horizontal dashed-lines in subplots~(d)--(f) correspond to $\lambda_{\rm max} =0$. Both measures ($l$ and $\lambda_{\rm max}$) yield the same conclusion.}}
	\label{fig:sl_ocs}
\end{figure}
We adopt the initial conditions $(0.10, 0.01, 0.02, 0.10)$ for the numerical simulation of Eq.~\ref{eq:sl}. Using the CCS (i.e., $\theta = 1$ in Eq.~\ref{eq:ocs}) with the parameter values $\omega_1 = 4$, $\omega_2 = 8$, and $\tau = 0$, the AD region is observed within the range $[\alpha_1, \alpha_2]$, i.e., $\alpha_1 \leq \alpha \leq \alpha_2$, where $\alpha_1 = 0.5$ and $\alpha_2 = 1.3$. Note that $\tau = 0$ implies that two SL oscillators are coupled through the diffusive coupling with twice the coupling strength (i.e., $2\alpha$). However, in this paper, our main focus is to deal with coupled oscillators in the presence of time-delay (i.e., $\tau > 0$). The bifurcation diagram of coupled SL oscillators at $\tau = 0.1$ is depicted in Fig.~\ref{fig:sl_dyn}a. In order to get Fig.~\ref{fig:sl_dyn}a, we plot the root mean square (rms) values of the variable $x'_1(t)$ (where $x'_1(t) = x_1(t) - \left\langle x_1(t) \right\rangle$, $\left\langle \cdot \right\rangle$ is the standard algebraic mean) after removing the initial $90 \%$ data as transient for a particular value of $\alpha$. The dashed-line and plus-marker ($+$) have been used to show the forward and backward variations of $\alpha$, respectively. Figure~\ref{fig:sl_dyn}b infers that individual SL oscillator exhibits limit cycle oscillations (LCO) in the regions $\alpha < \alpha_1$ and $\alpha > \alpha_2$. Thus, at $\alpha = \alpha_1$ and $\alpha = \alpha_2$, we observe transitions between a fixed point and LCO as we vary $\alpha$ monotonically --- implying the occurrence of Hopf bifurcation~\cite{strogatz07}. More explicitly, as the bifurcations are without exhibiting any hysteresis, these kinds of bifurcations are called supercritical Hopf bifurcations. In passing, plotting a bifurcation diagram using either of the three variables ($x_2,$ $ y_1,$ and $y_2$) yields the same conclusion.
Now, we are interested in employing the on-off coupling scheme (Eq.~\ref{eq:ocs}) to couple SL oscillators. Hence, we first have to choose an appropriate combination of scheme parameters $T$ and $\theta$. The average inter-peak separation of the SL oscillators is $1.57$ with $\alpha = 0$. Thus, following the rule of thumb~\cite{chen10,ghosh18}, we may choose any value  for $T$ within the range $[1, 10)$, and without loosing generality, for coupled SL oscillators, we choose $T$ as $1$ and $\theta$ arbitrarily. Here, we are interested in observing the effect of OCS in coupled SL oscillators. Our numerical experiment starts with $\tau = 0$, and the AD region enhances along the $\alpha$-axis after employing the OCS (Fig.~\ref{fig:sl_ocs}a). We can clearly observe in Fig.~\ref{fig:sl_ocs}b that as $\theta$ decreases, the width of the AD region (i.e., $\alpha_2 - \alpha_1$) increases. Similar results are obtained with a non-zero value of $\tau$. The red dashed-line plot in Fig.~\ref{fig:sl_ocs}c depicts the enhancement of AD region in the presence of time delay. It is apparent that \emph{the enhancement of the AD region along $\alpha$-axis is more significant in the presence of delay coupling}. 
\ag{The calculation of the maximum Lyapunov exponent ($\lambda_{\rm max}$) of coupled oscillators is also a suitable measure to confirm the AD state~\cite{saxena12}. A negative value of $\lambda_{\rm max}$ implies the AD state. For coupled SL oscillators, we have calculated $\lambda_{\rm max}$ at different values of $\alpha$. Figures~\ref{fig:sl_ocs}d and \ref{fig:sl_ocs}e correspond to $\tau = 0$. In the absence of OCS (i.e., $\theta = 1$ in Eq.~\ref{eq:ocs}), $\lambda_{\rm max}$ is negative within the range $[0.5, \, 1.3]$, and thus, we get AD regions withing the aforementioned range of $\alpha$ (Fig.~\ref{fig:sl_ocs}d). After the employment of OCS, at $\theta=0.5$, the region within which $\lambda_{\rm max}$ remains negative enhances to $[1.1,\, 6.5]$ (Fig.~\ref{fig:sl_ocs}e). Hence, we observe an enhancement of AD region along $\alpha$-axis in the presence of OCS. Finally, we make the time delay $\tau$ non-zero in Fig.~\ref{fig:sl_ocs}f with $\theta=0.5$, and compare the effectiveness of OCS in enhancing the AD regions with that while $\tau = 0$ (Fig.~\ref{fig:sl_ocs}e). We have seen that OCS is more effective in the presence of time delay. Therefore, in conclusion, using two different measures we have established that OCS is beneficial in enhancing the AD regions. In the rest of this paper, without losing generality, we have adopted the first measure (Eq.~\ref{eq:l}) to confirm the AD region.}
\ag{Now, we study this enhancement of the AD region in the presence of the OCS using a local stability analysis and try to determine the reason for this enhancement analytically. The uncoupled SL oscillator has a stable fixed point at $Z_j = 0$. We can linearize Eq.~\ref{eq:sl} around $Z_j = 0$ for the condition $T_s \gg T$ ($T_s$ is the system time-scale), and the corresponding characteristic equation is given by~\cite{reddy98,lakshmanan11}:
\begin{equation}
	\label{eq:char1}
	\text{det}(J - \lambda I) = 0,
\end{equation}
where $\lambda$ is the eigenvalue, $I$ is the identity matrix, and $J$ is the Jacobian. The explicit form of $J$ is as follows: 
\begin{equation}
\label{eq:jaco}
J = \begin{bmatrix}
a+i\omega_1 & K e^{-\lambda \tau } \\
K e^{-\lambda \tau } & a+i\omega_2
\end{bmatrix},
\end{equation}
where $a = (1-2K)$ and $K = \alpha \theta$. Using this explicit form of $J$, Eq.~\ref{eq:char1} can be rewritten as:
\begin{equation}
	\label{eq:char2}
	\lambda^2 - 2(a + i \omega) + (b_1 + i b_2) - K^2 e^{-2\lambda \tau} = 0.	
\end{equation}
In Eq.~\ref{eq:char2}, $b_1 = (a^2 - \omega^2 + \Delta^2/4)$, $b_2 = 2a\omega$, $\Delta = |\omega_1 - \omega_2|$, and $\omega = (\omega_1 + \omega_2)/2$. It (Eq.~\ref{eq:char2}) is a transcendental equation with infinite roots, and here we are interested in studying how the eigenvalues change in the parametric space $(\alpha, \Delta)$. Generally, eigenvalue $\lambda = (u + iv)$ is a complex number, and a negative value of $u$ infers the AD regions. In order to get the boundaries of the AD regions, we prescribe $u = 0$, i.e., $\lambda = iv$ in Eq.~\ref{eq:char2}, and then after separating the coefficients of real and imaginary parts, we get:
\begin{subequations}
\label{eq:sepa}
\begin{eqnarray}
	(v - \omega)^2 - \frac{\Delta^2}{4} - a^2 + K^2 \cos (2 v\tau)&=& 0,\label{eq:real}\\
	2a(v-\omega) - K^2 \sin(2v\tau) &=& 0 \label{eq:imag}.
\end{eqnarray}
\end{subequations}
}
\begin{figure}[htbp!]
	\hspace*{-2 mm}
	\includegraphics[width=40cm,height=10cm, keepaspectratio]{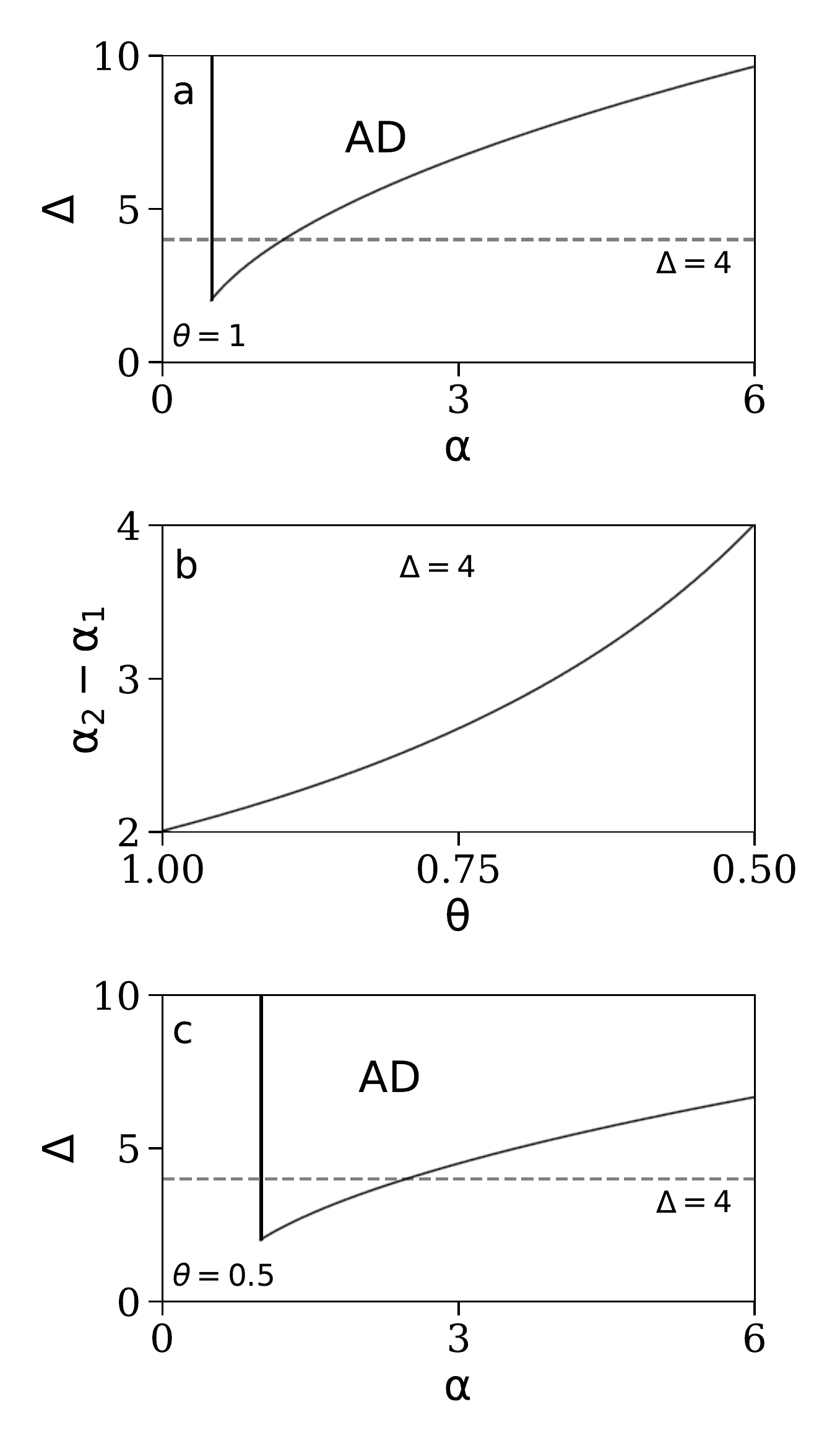}
	\caption{\ag{Bifurcation diagrams of coupled SL oscillator at $\tau = 0$. (a) The boundary of the AD region in the $\alpha$--$\Delta$ plane at $\theta = 1$. (b) The width of the AD region (Eq.~\ref{eq:diff}) is plotted as a function of $\theta$ at $\Delta = 4$. (c) The boundary of the AD region in the $\alpha$--$\Delta$ plane at $\theta = 0.5$. The horizontal dashed-lines in subplots (a) and (c) correspond to $\Delta = 4$.}}
	\label{fig:tau0}
\end{figure}
\ag{First, we consider that the time delay is absent~\cite{aronson90} (i.e., $\tau = 0$ in Eq.~\ref{eq:sl}). Substituting $\tau = 0$ in Eq.~\ref{eq:imag}, we get the conditions $\alpha = 1/2\theta$ and $v = \omega$. Furthermore, substituting $\tau = 0$ and $v = \omega$ in Eq.~\ref{eq:real}, we get $\alpha = (1/2\theta)(1 + \Delta^2/4)$. The explicit form of the eigenvalue from Eq.~\ref{eq:char2} for $\tau = 0$ is given by:
\begin{equation}
	\lambda = 1-2\alpha \theta \pm \sqrt{\alpha^2 \theta^2 - \frac{\Delta^2}{4}} + i \omega.
\end{equation}
On a monotonically increase in $\alpha$ from $0$, when $\Delta > 2$, the Hopf bifurcation occurs at $\alpha = 1/2\theta$. For $\Delta > 2$, as $\alpha > 1/2\theta$, a pair of eigenvalues with negative real parts are generated --- implying the AD region. On further increase in $\alpha$, this AD region continues existing up to $\alpha = (1/2\theta)(1 + \Delta^2/4)$. In contrast, no AD region is detected for $\Delta < 2$. Hence, for the example in hand, this AD region is bounded by the curves $\alpha_1 = 1/2\theta$ and $\alpha_2 = (1/2\theta)(1 + \Delta^2/4)$ in the $\alpha$--$\Delta$ plane with the condition $\Delta \geq 2$. In Fig.~\ref{fig:tau0}, we have plotted the boundaries of the AD region in the $\alpha$--$\Delta$ plane for two different values of $\theta$. We recall that $\theta = 1$ corresponds to the CCS, and the AD region using the CCS is depicted in Fig.~\ref{fig:tau0}a. The width of the AD region along $\alpha$-axis is given by:
\begin{equation}
\label{eq:diff}
\alpha_2 - \alpha_1 = \frac{\Delta^2}{8 \theta}.
\end{equation}
This width (Eq.~\ref{eq:diff}) is plotted as a function of $\theta$ for $\Delta = 4$ in Fig.~\ref{fig:tau0}b. It is clearly visible in Fig.~\ref{fig:tau0}b that with the decrease in $\theta$, $\alpha_2 - \alpha_1$ increases. Finally, for $\theta = 0.5$, the AD region is depicted in Fig.~\ref{fig:tau0}c. In Figs.~\ref{fig:tau0}a and \ref{fig:tau0}c, the horizontal dashed lines correspond to $\Delta = 4$. Thus, analytically, we have established the enhancement of the AD region in the presence of OCS.} 
\begin{figure}[htbp!]
	\hspace*{-2 mm}
	\includegraphics[width=20cm,height=3cm, keepaspectratio]{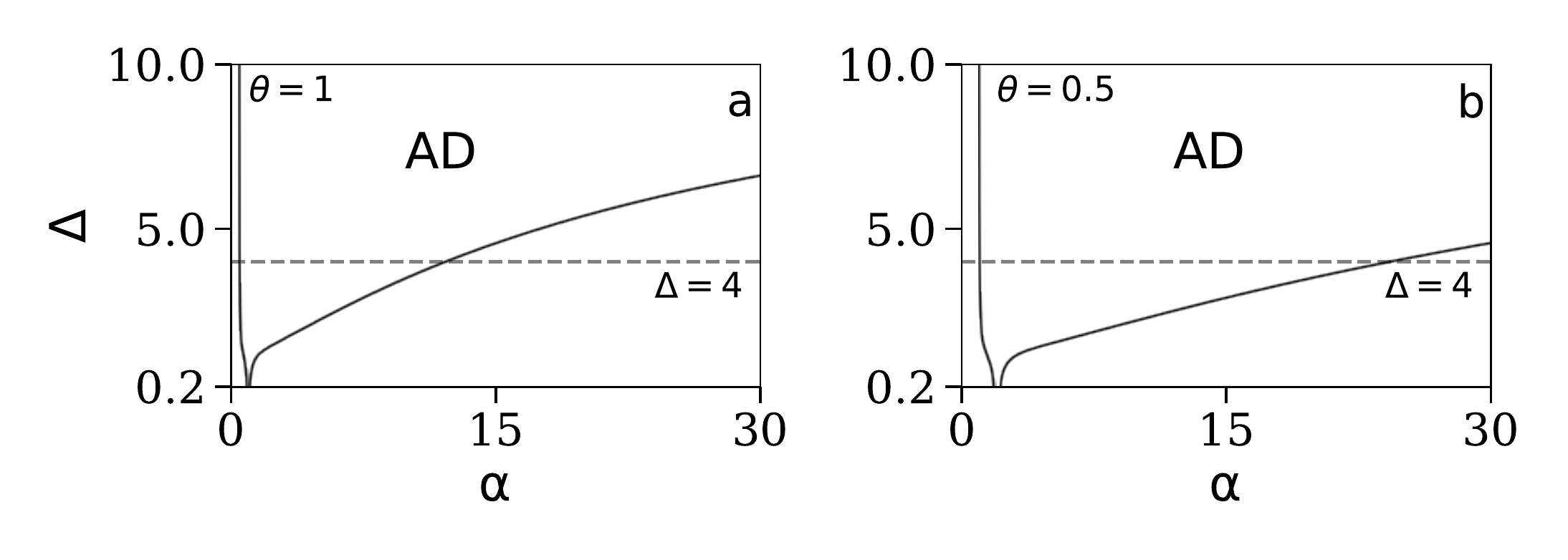}
	\caption{\ag{Bifurcation diagrams of coupled SL oscillator at $\tau = 0.08$ and $\omega = 6$. It is clearly visible that for a fixed value of $\Delta$, the width of the AD region along $\alpha$-axis is larger in subplot~(b) compared to that in subplot~(a). Thus, with the decrease in $\theta$, an enhancement of the AD region is depicted. The horizontal dashed-lines in both subplots correspond to $\Delta = 4$.} }
	\label{fig:tau1}
\end{figure}
\ag{Next, we make the time delay $\tau$ non-zero and study Eq.~\ref{eq:sl}. In order to get the boundary of the AD region, we introduce a new function $F(v)$ as follows:
\begin{equation}
\label{eq:f}
F(v) := \frac{(v-\omega)}{\sin(2v\tau)}.	
\end{equation}
Here, $F(v)$ is a piecewise continuous function of $v$ with singularities at $v_n = n\pi/2\tau$, where $n$ is an integer. In terms of $F(v)$, Eq.~\ref{eq:sepa} can be rewritten as:
\begin{subequations}
	\label{eq:sepa_tau}
	\begin{eqnarray}
	\alpha &=& \frac{-2F \pm \sqrt{2F(2F+1)}}{\theta},\label{eq:real_tau}\\
	\Delta^2 &=&  4 \alpha^2 \theta^2 \cos (2 v\tau) + 4(v - \omega)^2 \nonumber \\ && -4(1-2\alpha \theta)^2 \label{eq:imag_tau}.
	\end{eqnarray}
\end{subequations}
In order to calculate $\alpha$ and $\Delta$ from Eq.~\ref{eq:sepa_tau}, we adopt $v$ from the open intervals $(v_n, v_{n+1})$ and other two parameters $\tau$ and $\omega$ are chosen as fixed. We obtain two different values of $\alpha$ from Eq.~\ref{eq:real_tau}, and let $\alpha_+$ ($\alpha_-$) be the corresponding value of $\alpha$ due to the positive (negative) sign in Eq.~\ref{eq:real_tau}. Consequently, two curves $C_{+} := C_{+}(\alpha_+, \Delta)$ and $C_{-} := C_{-}(\alpha_-, \Delta)$ construct the boundary of the AD region in the $\alpha$--$\Delta$ plane. We have chosen $\omega_1 = 4$ and $\omega_2 = 8$; hence, the average frequency $\omega = 6$. Fig.~\ref{fig:tau1} depicts the AD regions for $\tau = 0.08$ in the absence ($\theta = 1.0$) and presence ($\theta = 0.5$) of OCS. We can conclude that, similar to Fig.~\ref{fig:tau0}, OCS is enhancing the AD region along $\alpha$-axis for non-zero value of $\tau$. To this end, we mention that the width of the AD region along $\alpha$-axis is larger in the presence of time delay.}
\begin{figure}[htbp!]
	\hspace*{-1 mm}
	\includegraphics[width=50cm,height=5.7cm, keepaspectratio]{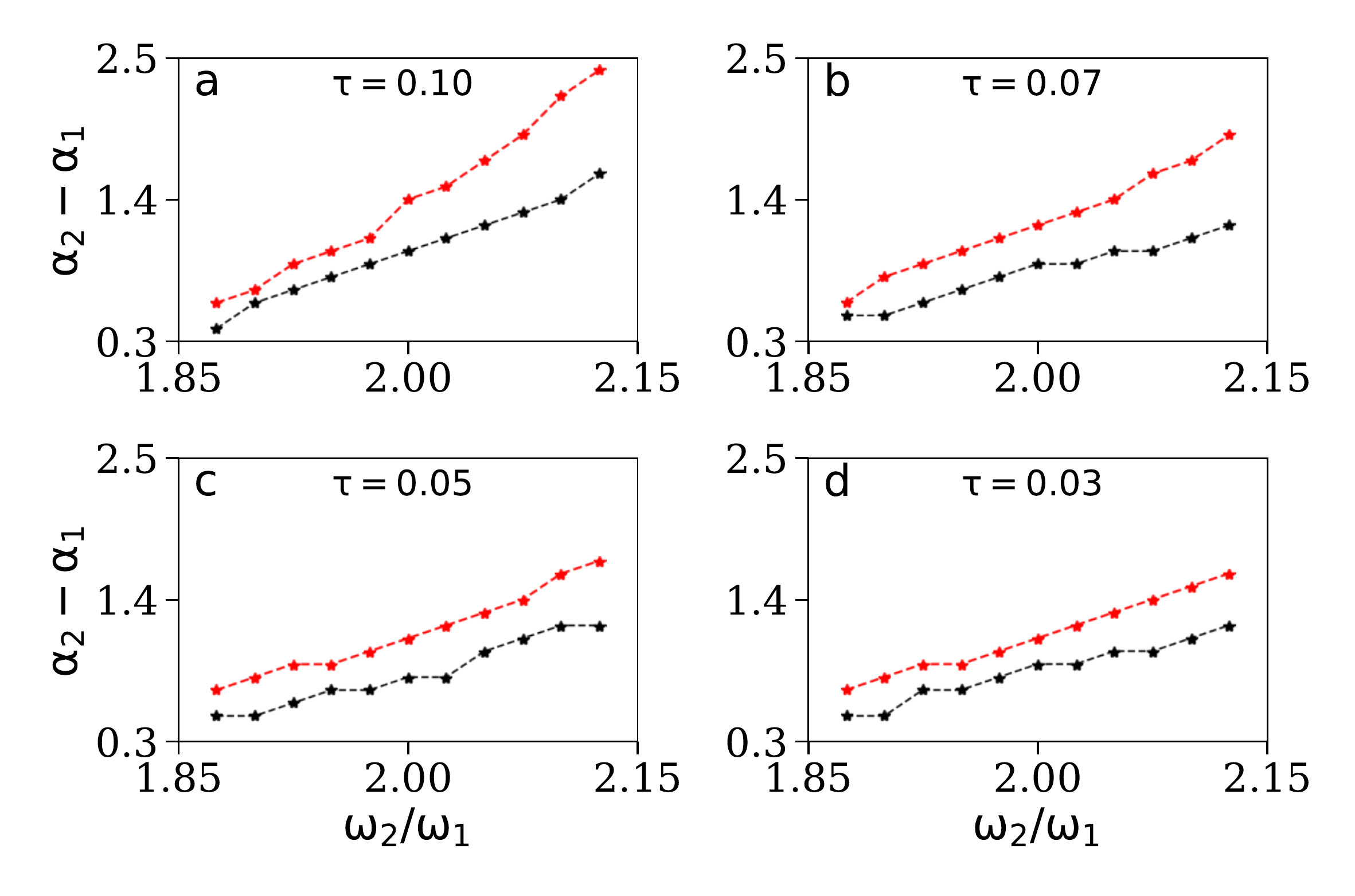}
	\caption{(\textit{Color online}) The effect of OCS is studied in coupled SL oscillators (Eq.~\ref{eq:sl}) by varying the frequency ratio $\omega_2/\omega_1$ at a fixed value of $\tau$. Here, the red and the black plots correspond to OCS and CCS, respectively. In all four cases, the width of the AD regions ($\alpha_2 - \alpha_1$) is larger using OCS than that using CCS. In all four cases, $\theta$ is chosen as $0.5$.} 
	\label{fig:freq_mis_sl}
\end{figure}
Thus, we have ascertained that the employment of OCS is fruitful to enhance the AD region along the coupling strength ($\alpha$) parameter  (Figs.~\ref{fig:sl_ocs}, \ref{fig:tau0}, and \ref{fig:tau1}). Additionally, experiments have shown the suitability of other system parameters such as time delay or frequency ratio as control parameter~\cite{pawar17,mondal17_chaos,raaj19,moon20}. Therefore, we use the frequency ratio ($\omega_2/\omega_1$) as a control parameter, and the effects of OCS using $\omega_2/\omega_1$ as the control parameter are depicted in Fig.~\ref{fig:freq_mis_sl}. The red and the black plots correspond to OCS and  CCS, respectively. We check the effect of OCS after varying the ratio $\omega_2/\omega_1$ at a fixed time delay $\tau$. More explicitly, we calculate the width of the AD regions ($\alpha_2 - \alpha_1$) for different values of $\omega_2/\omega_1$. Figure~\ref{fig:freq_mis_sl} supports that the width, $\alpha_2 - \alpha_1$, always has greater values for OCS than that in the case of CCS. Also, we observe that $\alpha_2 - \alpha_1$ increases with the increase in $\omega_2/\omega_1$, and this enhancement is observed for all four values of $\tau$. Henceforth, we use the same colour code in all figures: black for CCS and red for OCS, and markers: star ($*$) and cross-mark ($\times$) in plotting the width of the AD regions and the boundaries of the AD regions as a function of the control parameter, respectively.
\begin{figure}[htbp!]
	\hspace*{-1 mm}
	\includegraphics[width=50cm,height=5.7cm, keepaspectratio]{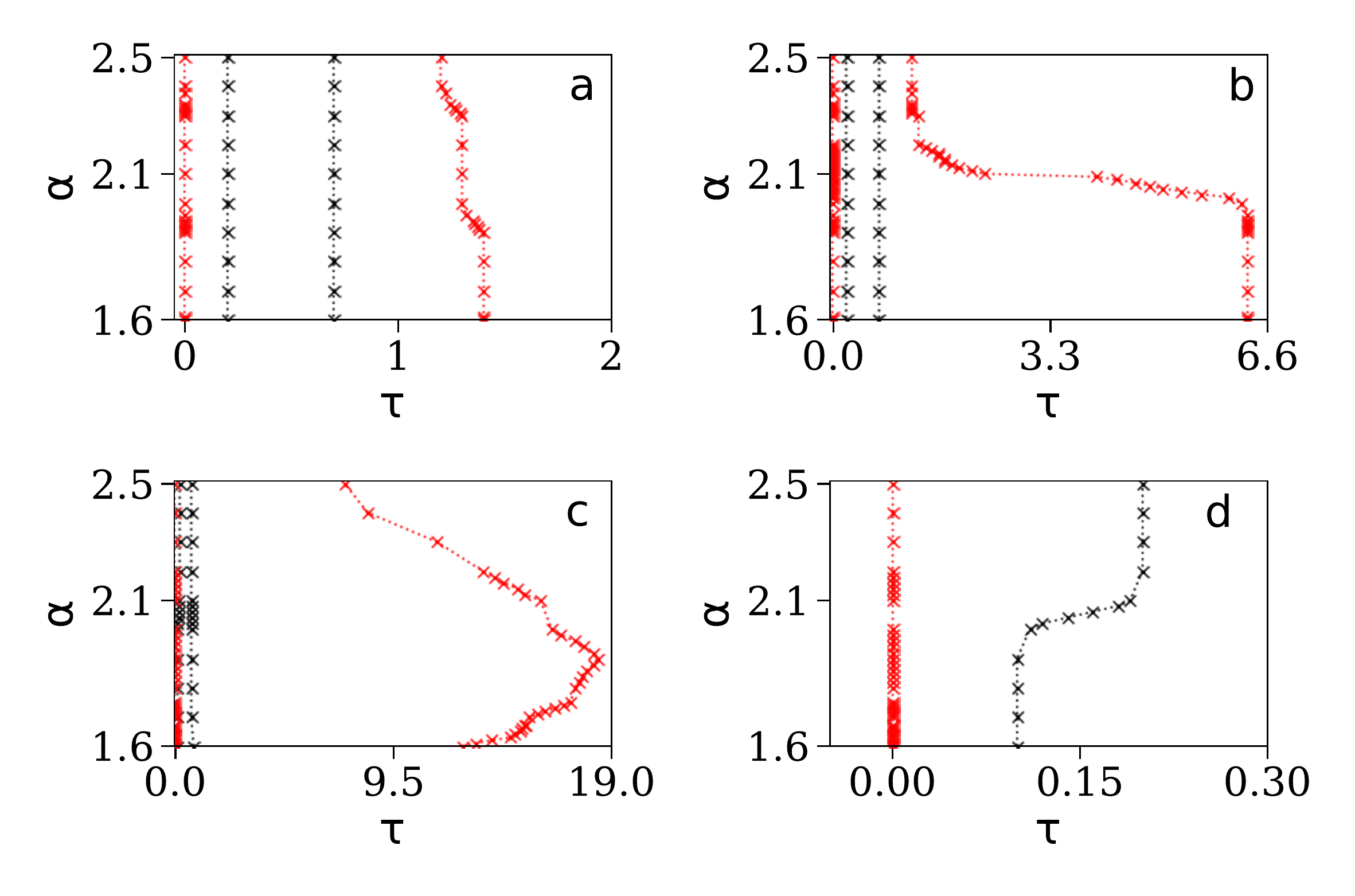}
	\caption{(\textit{Color online}) The effect of OCS is studied in coupled SL oscillators (Eq.~\ref{eq:sl}) by varying the time delay ($\tau$) at a fixed frequency ratio $\omega_2/\omega_1$. The parameter $\omega_2/\omega_1$ has the values $1.87,$ $2.00,$ and $2.12$ for the subplots (a), (b), and (c), respectively. Subplot~(d) is the zoom portion of subplot~(c) around $\tau = 0.15$. The on-off coupling scheme parameter (Eq.~\ref{eq:ocs}) $\theta$ is chosen as $0.5$. In all three cases, the width of the AD regions along $\tau$-axis are larger using OCS than that using CCS.}
	\label{fig:sl_time_delay}
\end{figure}
Next, we extend our study to focus on the effect of the third system parameter $\tau$ in Eq.~\ref{eq:sl}. Here, we vary $\tau$ monotonically, keeping the frequency ratio ($\omega_2/\omega_1$) unaltered, and the corresponding results are depicted in Fig.~\ref{fig:sl_time_delay}. The enhancement of AD regions along $\tau$-axis is clearly depicted for different fixed values of $\omega_2/\omega_1$ in Fig.~\ref{fig:sl_time_delay}, i.e., the AD regions extend along the $\tau$-axis in all cases. Note that, even with $\tau = 0$, we achieve AD after employing the OCS in coupled SL oscillators (Fig.~\ref{fig:sl_time_delay}). 
Thus, we have studied the effect of OCS using three control parameters in coupled SL oscillators. In all three cases, the employment of OCS is worthwhile in enhancing the AD regions along the control parameter axis. However, as the time delay and frequency ratio  is more suitable  control parameter in experiments~\cite{pawar17,mondal17_chaos,manoj18,raaj19,moon20}, from now onward, we use these two parameters, time delay and frequency ratio, as the control parameters to study the effect of OCS on AD.     
\subsection{Coupled R\"ossler oscillators}
\label{sec:ross}
\begin{figure}[htbp!]
	\hspace*{-1 mm}
	\includegraphics[width=50cm,height=5.7cm, keepaspectratio]{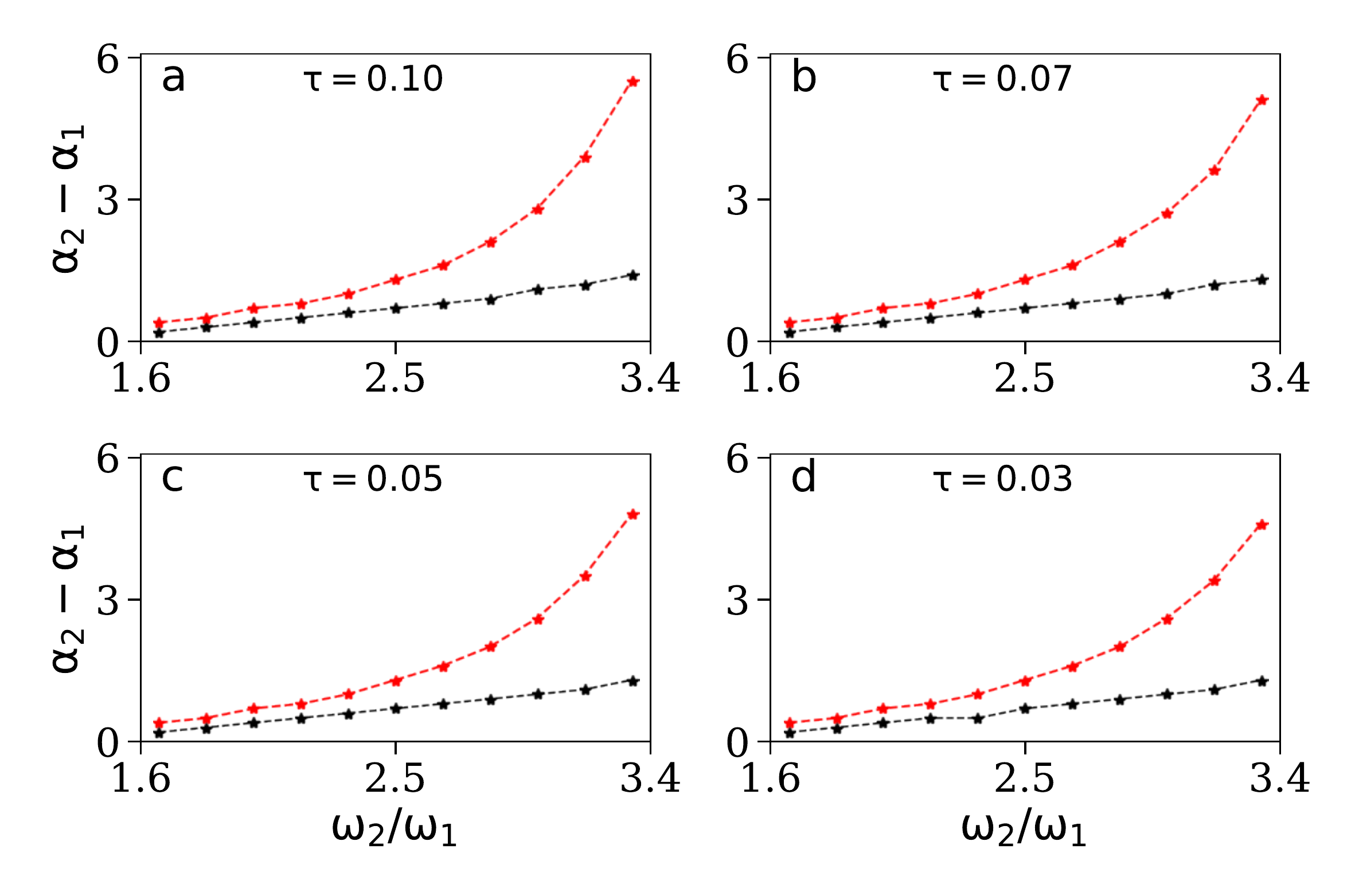}
	\caption{(\textit{Color online}) The effect of OCS is studied in coupled R\"ossler oscillators (Eq.~\ref{eq:ross}) as a function of frequency ratio $\omega_2/\omega_1$ at fixed $\tau$. In all four cases, the width of the AD regions ($\alpha_2 - \alpha_1$) are greater using OCS than that using CCS.}
	\label{fig:freq_mis_ross}
\end{figure}

\begin{figure}[htbp!]
	\hspace*{1 cm}
	\includegraphics[width=80cm,height=9.5cm, keepaspectratio]{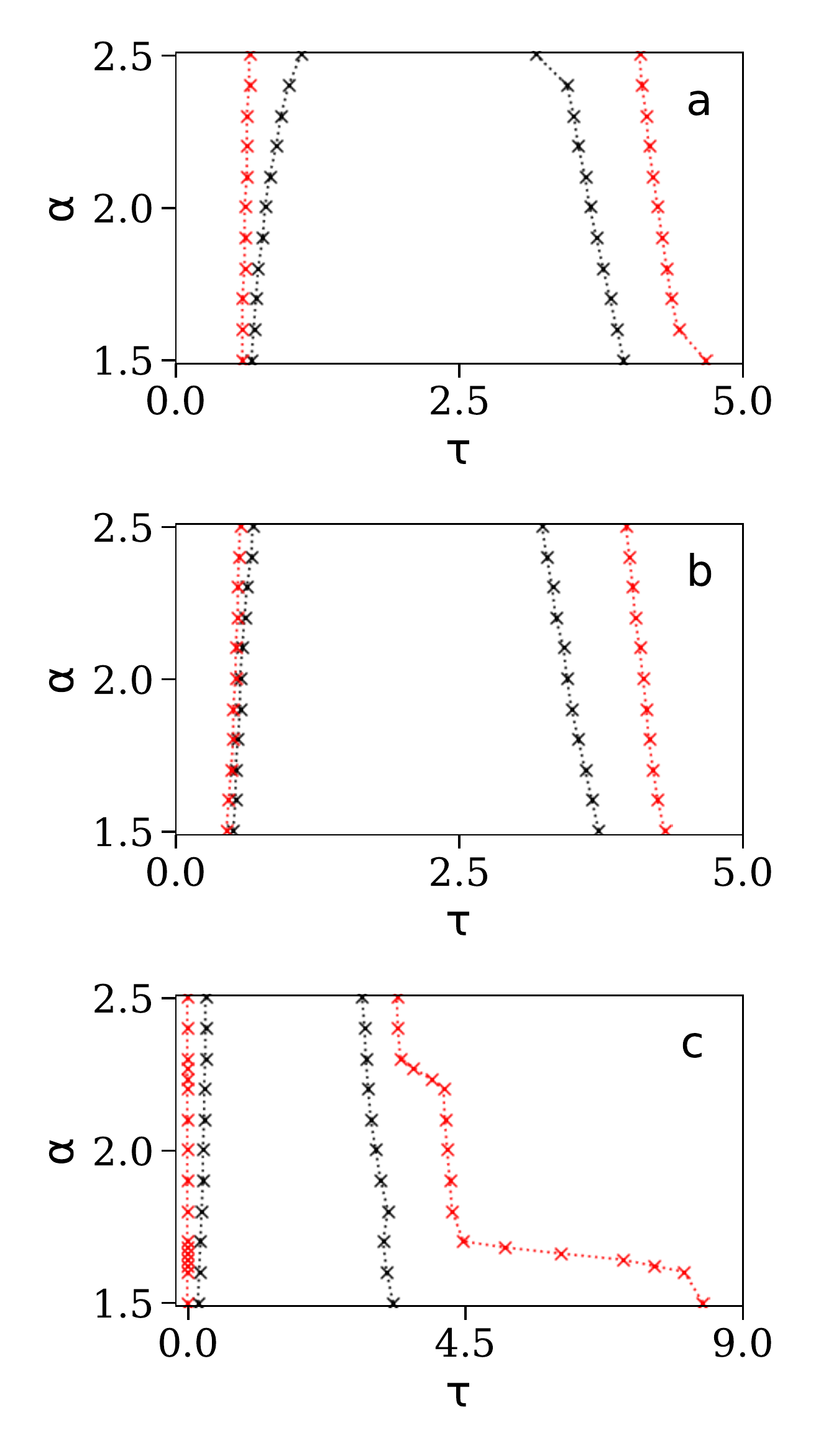}
	\caption{(\textit{Color online}) The effect of OCS is studied in coupled R\"ossler oscillators (Eq.~\ref{eq:ross}) by varying the time delay ($\tau$) at three fixed frequency ratio $\omega_2/\omega_1$. The parameter $\omega_2/ \omega_1$ has the values $2.0,$ $2.3,$ and $3.3$ for the subplots (a), (b), and (c), respectively. In all three cases, the AD region using CCS is a fraction of that using OCS.}
	\label{fig:ross_time_delay}
\end{figure}

Now, we switch to the second example of this section: R\"ossler oscillator~\cite{roessler76}. It is a three dimensional, autonomous, chaotic oscillator. Thus, for the coupled R\"ossler oscillators, following Eq.~\ref{eq:1_ocs}, $\mathbf{x}_1 = (x_1, y_1, z_1)$ and $\mathbf{x}_2 = (x_2, y_2, z_2)$. The explicit form of the equations of motion are as follow:
\begin{subequations}
	\label{eq:ross}
	\begin{eqnarray}
	\frac{d{x}_j}{dt} =&& -\omega_j (y_j + z_j) + \alpha_d \cdot \chi (t) \cdot  (x_l - x_j) \nonumber \\ &&+ \alpha_{\tau} \cdot \chi (t) \cdot  (x_{l \tau} - x_j),\label{eq:ross_a}\\
	\frac{d{y}_j}{dt} =&& \omega_j (x_j + 0.15 y_j) + \alpha_d \cdot \chi (t) \cdot  (y_l - y_j) \nonumber \\ &&+ \alpha_{\tau} \cdot \chi (t) \cdot  (y_{l \tau} - y_j),\label{eq:ross_b}\\
	\frac{d{z}_j}{dt} =&& \omega_j \left(0.4 + z_j (x_j - 8.5)\right) \nonumber + \alpha_d \cdot \chi (t) \cdot  (z_l - z_j)  \\ &&+ \alpha_{\tau} \cdot \chi (t) \cdot  (z_{l \tau} - z_j),\label{eq:ross_c}
	\end{eqnarray}
\end{subequations} 
where $l = 1, 2$ with $l \neq j$. We choose the natural frequencies ($\omega_j$) as $0.6$ and $1.4$, respectively~\cite{sun18}. Similar to the previous example, we choose $\alpha_{\tau} = \alpha_d = \alpha$, and the initial condition is adopted as $(-9, 0, 0, -9.01, 0.01, 0)$.
\begin{figure}[htbp!]
	\hspace*{-0.0 cm}
	\includegraphics[width=70cm,height=9.3 cm, keepaspectratio]{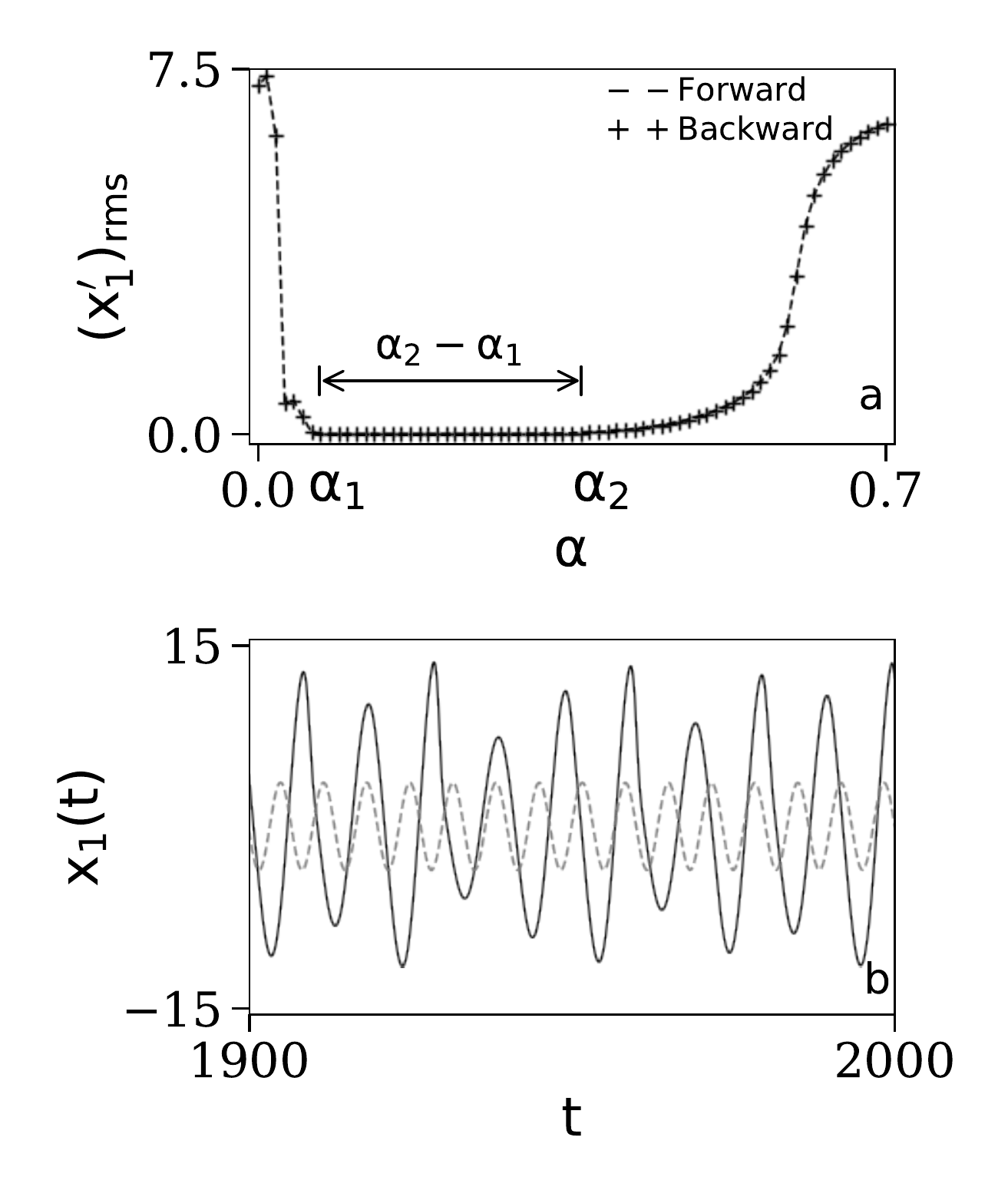}
	\caption{Transitions chaotic dynamics -- AD -- periodic dynamics are detected as we increase $\alpha$ monotonically. The time delay ($\tau$) and natural frequencies ($\omega_1$ and $\omega_2$) remain fixed in both subplots at $\tau = 0.1$, $\omega_1 = 0.6$, and $\omega_2 = 1.4$. (a) Bifurcation diagram of the coupled R\"ossler oscillators is plotted using the variable $x_1(t)$. (b) The variable $x_1(t)$ is plotted for two different values of $\alpha$: $0.02$ (black line plot) and $0.6$ (gray dashed-line plot). Chaotic and periodic dynamics are ascertained at $\alpha = 0.02$ and $0.6$, respectively.}
	\label{fig:ross_dyn}
\end{figure}
The isolated R\"ossler oscillator, using the parameter values mentioned in Eq.~\ref{eq:ross}, exhibits chaotic dynamics. Thus, for this example, we study a transition from the chaotic dynamics to a fixed point as the coupling strength (as a control parameter) is increased monotonically. We use Fig.~\ref{fig:ross_dyn} to understand this transition more clearly. The bifurcation diagram (Fig.~\ref{fig:ross_dyn}a) for coupled R\"ossler oscillators is drawn following the algorithm discussed in Sec.~\ref{sec:sl}. The existence of chaotic dynamics at lower values of the control parameter is depicted in Fig.~\ref{fig:ross_dyn}b. The transition from chaotic dynamics to AD is ascertained as we increase the control parameter monotonically. We observe the periodic dynamics on further increase in the control parameter. Prasad~\cite{prasad05} had already discussed in detail the route from an oscillatory state to the AD in delay-coupled oscillators. The examples of both SL and R\"ossler oscillators had been studied there.
For the R\"ossler oscillators (Eq.~\ref{eq:ross}) with $\alpha = 0$, the average inter-peak separation is $4.33$. Thus, we may choose any value of $T$ within the range $[1, 10)$, and similar to the previous example, we choose $(T, \theta) = (1, 0.5)$ to employ the on-off coupling (Eq.~\ref{eq:ocs}) in coupled R\"ossler oscillators (Eq.~\ref{eq:ross}), and the results are depicted in Figs.~\ref{fig:freq_mis_ross} and \ref{fig:ross_time_delay}.  
%
%
%
%
In order to study the effect of OCS on coupled R\"ossler oscillators, we begin with the frequency ratio as the control parameter at a fixed value of time delay (Fig.~\ref{fig:freq_mis_ross}). The width of the AD regions enhances with the increase in $\omega_2/\omega_1$ in the presence of OCS, whereas using CCS, the enhancement is comparatively small. Also, we mention that the width of the AD region increases monotonically as the control parameter $\omega_2/\omega_1$ increases. We obtain the same conclusions for four different values of $\tau$. Besides, we use the time delay $\tau$ as the control parameter at fixed values of $\omega_2/\omega_1$ in Fig.~\ref{fig:ross_time_delay}. The corresponding results are shown for three different values of $\omega_2/\omega_1$. The AD region using CCS (region within the black lines) is a fraction of the AD region using OCS (region within the red lines). Therefore, the effectiveness of OCS to enhance the AD regions is evident in all three cases of Fig.~\ref{fig:ross_time_delay}.
To summarize, we have taken two low-dimensional mathematical models of coupled oscillators to study the effectiveness of OCS. In both the examples, we have obtained an enhancement of AD region along the parameter axis with OCS. Having established that fact, in Sec.~\ref{sec:rijke}, we adopt a mathematical model representing a thermoacoustic system, the horizontal Rijke tube. 
\subsection{Model of coupled horizontal Rijke tubes}
\label{sec:rijke}
\begin{figure}[htbp!]
	\hspace*{0.1 cm}
	\includegraphics[width=50cm,height=5.3cm, keepaspectratio]{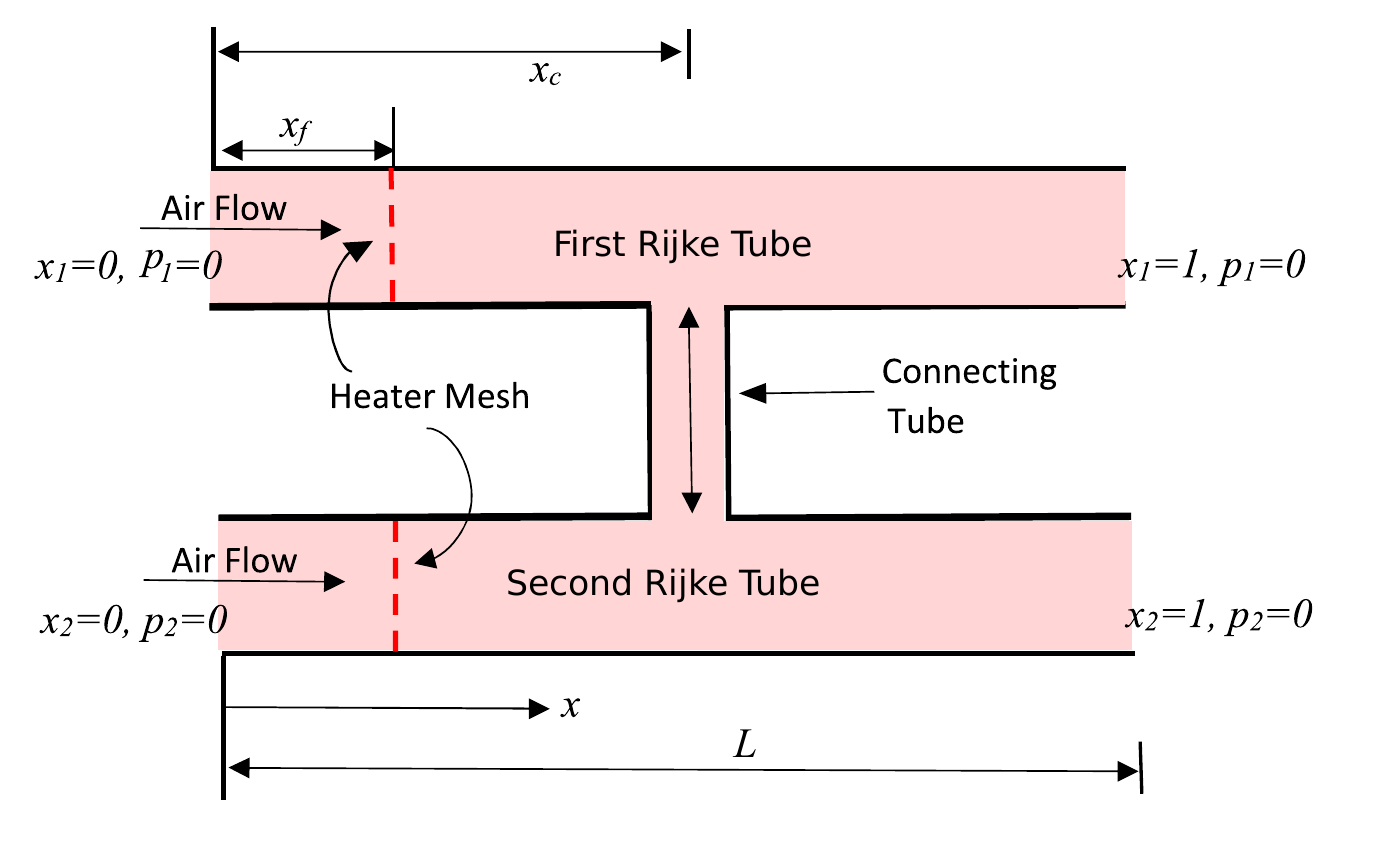}
	\caption{The schematic diagram of coupled horizontal Rijke tubes~\cite{rijke59}. Both the Rijke tubes have same length $L$ and are connected through a connecting tube. The spatial distance $x$ is normalized by the duct length, i.e., $x := x/L$. The connecting tube is situated at the spatial distance $x_c = 0.5$.} 
	\label{fig:rijke_schematic}
\end{figure}
In this study, we adopt the example of coupled horizontal Rijke tubes --- each tube has a cuboid duct with two ends open, and an electrically heated wire-mesh is placed within the duct for heating the flow through it. The schematic diagram of coupled horizontal Rijke tubes is depicted in Fig.~\ref{fig:rijke_schematic}. A horizontal Rijke tube is one of the simplest thermoacoustic systems exhibiting an oscillatory instability, known as thermoacoustic instability~\cite{balasubramanian08,subramanian10,juniper18,sujith21}. {The positive feedback between heat release and acoustic field yield this instability~\cite{sujith21}.} An earlier study~\cite{thomas18} supports the existence of AD in the coupled horizontal Rijke tube model using time-delay and dissipative couplings. Here, we study the effect of OCS on AD using this model. In Appendix~\ref{sec:appen_rijke}, we present an elaborate discussion on the governing equations of a single, uncoupled horizontal Rijke tube and the chosen parameters values used for numerical analysis. The equations of motion of two Rijke tubes subjected to diffusive and time delay couplings (i.e., following Eq.~\ref{eq:1}) are as follow:
\begin{subequations}
	\label{eq:coupled_rijke}
	\begin{eqnarray}
	&&\frac{d \eta^j_k}{d t} = \dot{\eta}^j_k, \label{eq:coupled_rijke_a}\\
	&&\frac{d \dot{\eta}^j_k}{d t} + 2\zeta_k \omega_k \dot{\eta}^j_k + \omega^2_k {\eta}^j_k \nonumber\\ && = - k \pi K^j \left[ \sqrt{|{\frac{1}{3} + u_f(t-\tau_1)}|} - \sqrt{\frac{1}{3}}\right] \nonumber \sin(k \pi x_f) \label{eq:coupled_rijke_b}\nonumber \\ && + \alpha_d \cdot \left(\dot{\eta}^l_k - \dot{\eta}^j_k\right) + \alpha_{\tau} \cdot \left(\dot{\eta}^l_{k\tau} - \dot{\eta}^j_k\right). 
	\end{eqnarray}
\end{subequations}
We recall that $j = 1, 2$, $l = 1, 2$ with $l \neq j$, $k = 1, 2, \cdots, N$, and $\dot{\eta}^l_{k\tau} = \dot{\eta}^l_{k} (t - \tau)$. The corresponding governing equations of coupled Rijke tubes after incorporating OCS become:
\begin{subequations}
	\label{eq:coupled_rijke_ocs}
	\begin{eqnarray}
	&&\frac{d \eta^j_k}{d t} = \dot{\eta}^j_k, \label{eq:coupled_rijke_ocs_a}\\
	&&\frac{d \dot{\eta}^j_k}{d t} + 2\zeta_k \omega_k \dot{\eta}^j_k + \omega^2_k {\eta}^j_k \nonumber\\ && = - k \pi K^j \left[ \sqrt{|{\frac{1}{3} + u_f(t-\tau_1)}|} - \sqrt{\frac{1}{3}}\right] \nonumber \sin(k \pi x_f) \label{eq:coupled_rijke_ocs_b}\nonumber \\ && + \alpha_d \cdot \chi (t) \cdot \left(\dot{\eta}^l_k - \dot{\eta}^j_k\right) + \alpha_{\tau} \cdot \chi (t) \cdot \left(\dot{\eta}^l_{k\tau} - \dot{\eta}^j_k\right). 
	\end{eqnarray}
\end{subequations}
\begin{figure}[htbp!]
	\hspace*{1 cm}
	\includegraphics[width=65cm,height=7.8cm, keepaspectratio]{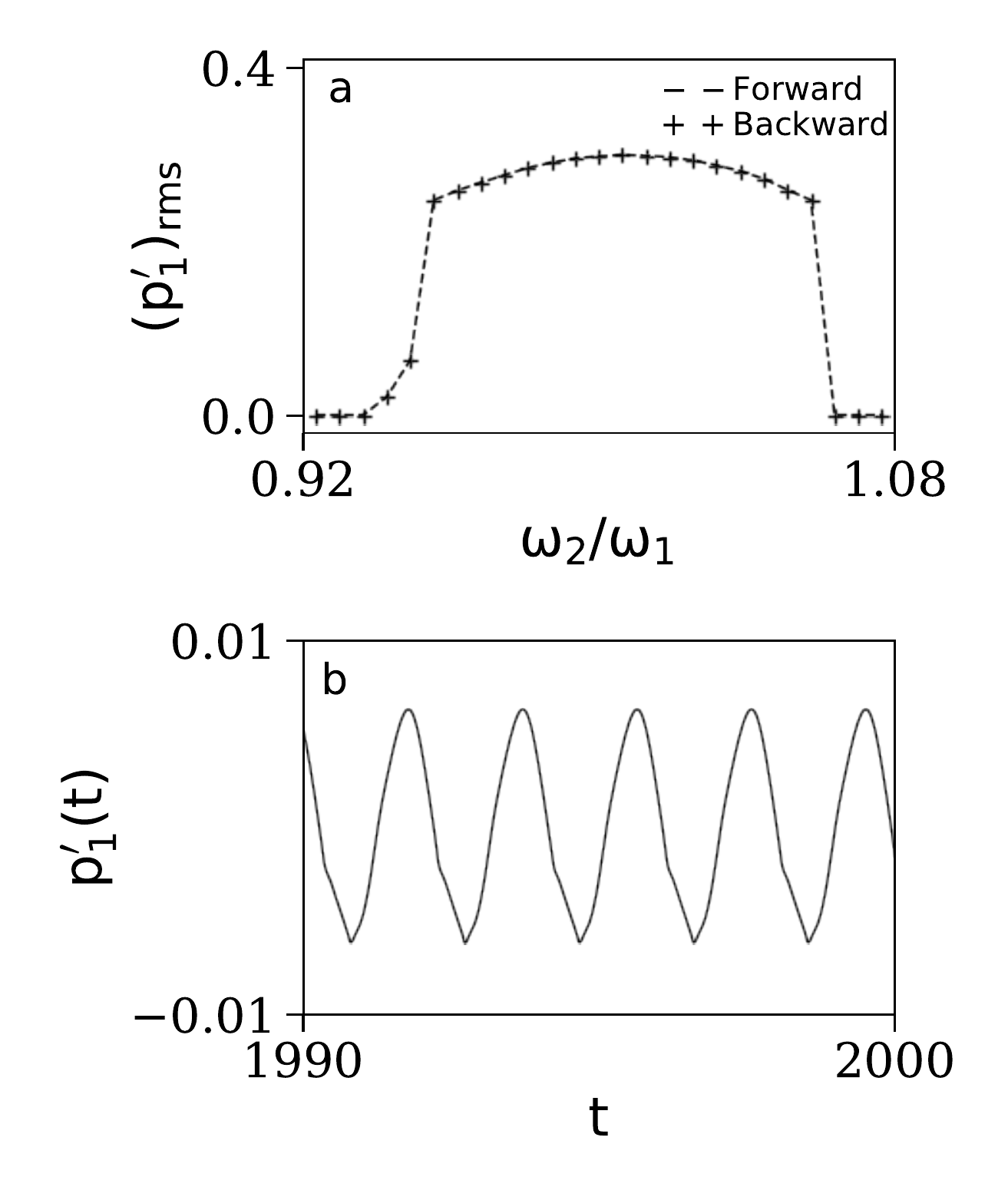}
	\caption{ (a) The transition AD -- LCO -- AD is observed in coupled horizontal Rijke tube model. (b) LCO is detected in the variation of $p_1(t)$ at $\omega_2/ \omega_1 = 1$. In both plots, rest parameter values are adopted as $\tau = 0.5$, $\alpha_d = 0.25$, and $\alpha_{\tau} = 0.05$.}
	\label{fig:rijke_dyna}
\end{figure}
For a physical system, we might not have the liberty to choose an identical value for  $\alpha_d$ and $\alpha_{\tau}$, therefore, in this model, we choose non-identical values of those coupling strength parameters to study AD~\cite{thomas18,dange19}. Thus, for the example in hand, we need to deal with four system parameters: time delay ($\tau$), frequency ratio ($\omega_2/\omega_1$), diffusion coupling strength ($\alpha_d$), and time-delay coupling strength ($\alpha_{\tau}$). Similar to the previous two examples, here, we study the transition between the oscillatory state and AD in coupled horizontal Rijke tubes model. We use the same initial conditions, as mentioned in Appendix~\ref{sec:appen_rijke}, for both the tubes. A bifurcation diagram (Fig.~\ref{fig:rijke_dyna}a) is drawn for parameter values $\tau = 0.5$, $\alpha_d = 0.25$, and $\alpha_{\tau} = 0.05$, and the transition AD -- LCO -- AD is observed. Figure~\ref{fig:rijke_dyna}b depicts the LCO of $p_1'(t)$ at $\omega_2/\omega_1 = 1$. Note that $(p_1')_{\rm rms}$ follow the same paths during the forward and backward variations of $\omega_2/\omega_1$, implying the occurrence of supercritical Hopf bifurcations.
\begin{figure}[htbp!]
	\hspace*{1 cm}
	\includegraphics[width=65cm,height=7.8cm, keepaspectratio]{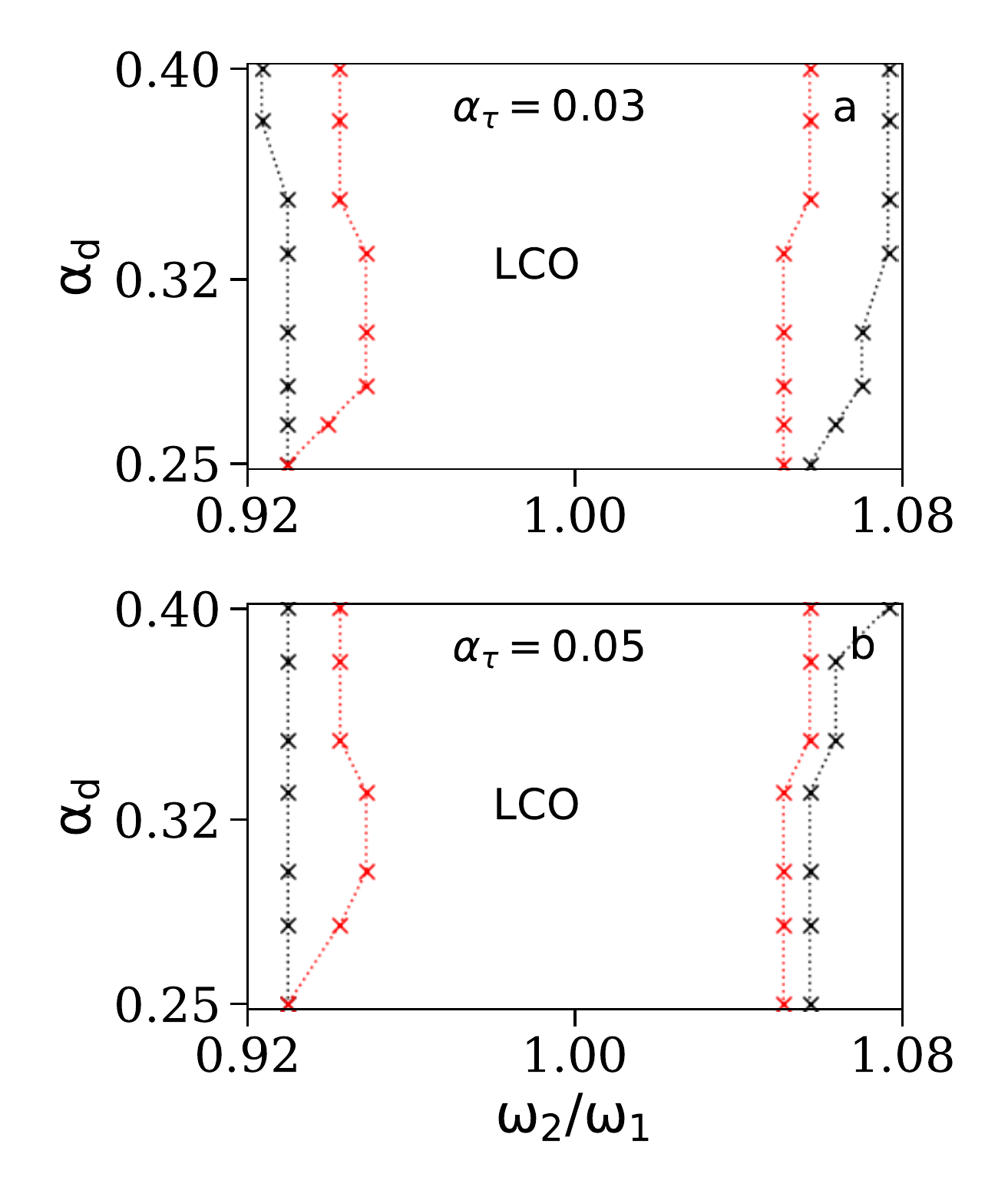}
	\caption{\textit{(Color online)} The effect of OCS is studied in coupled Rijke tubes (Eq.~\ref{eq:coupled_rijke_ocs}) as a function of the frequency ratio ($\omega_2/\omega_1$) at a fixed time delay $\tau = 0.5$. The width of the AD region along $\omega_2/\omega_1$-axis is larger using OCS compare to that of using CCS. }
	\label{fig:rijke_diff_kd}
\end{figure}
Furthermore, we choose the system parameter $K^1 = K^2 = 0.92$ (representing heater power) in Eq.~\ref{eq:coupled_rijke_ocs}~\cite{thomas18}. For the uncoupled Rijke tube, i.e., putting $\alpha_d = \alpha_{\tau} = 0$ in Eq.~\ref{eq:coupled_rijke_ocs}, the average inter-peak separation is $1.92$, and we choose $T = 2$  (and $\theta = 0.5$) to employ the on-off coupling (Eq.~\ref{eq:ocs}) in coupled Rijke tubes model (Eq.~\ref{eq:coupled_rijke_ocs}). 
\begin{figure}[h]
	\hspace*{1.0 cm}
	\includegraphics[width=75cm,height=7.8cm, keepaspectratio]{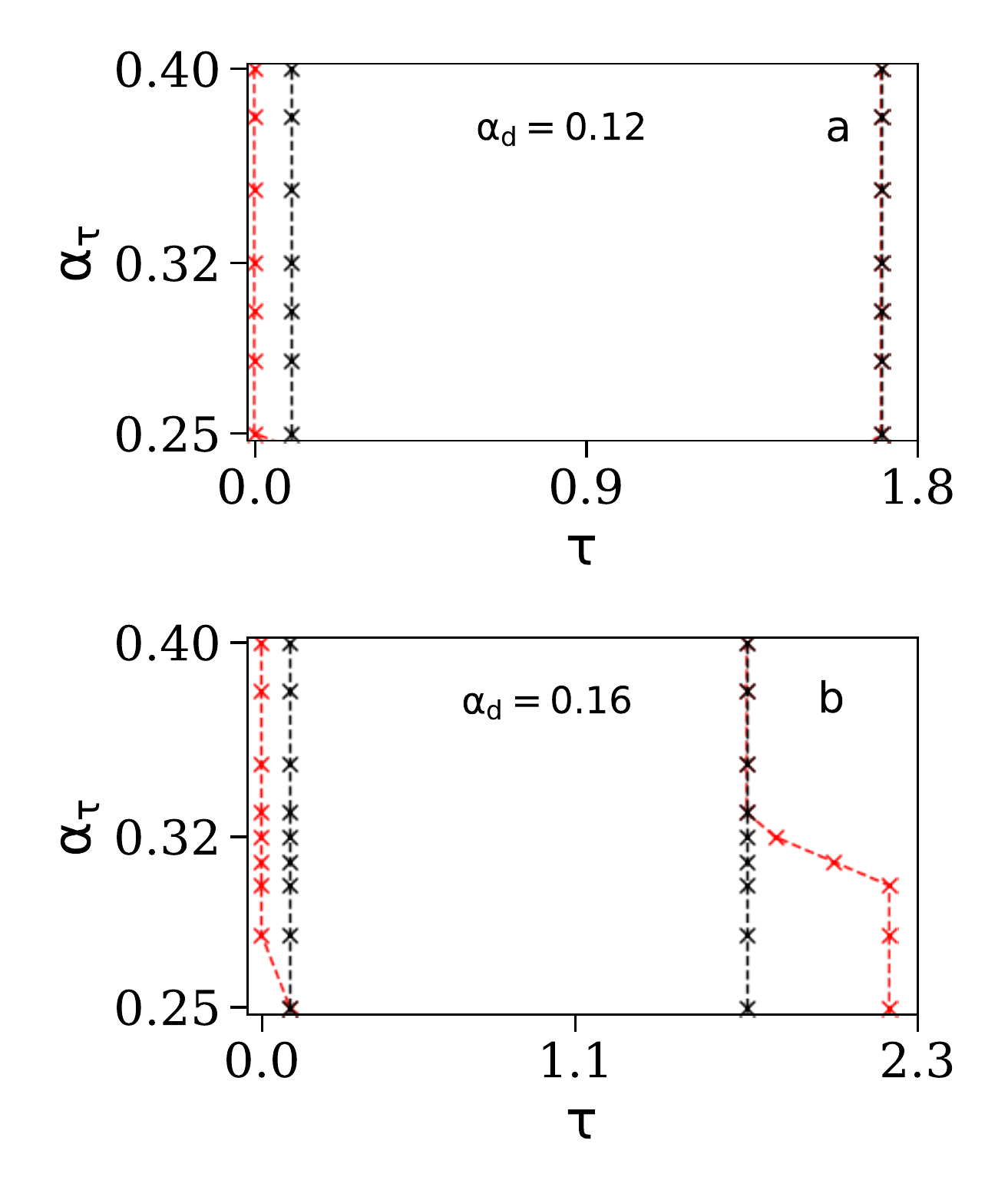}
	\caption{(\textit{Color online}) The effect of OCS is studied in coupled Rijke tubes (Eq.~\ref{eq:coupled_rijke_ocs}) by varying the time delay ($\tau$) at a fixed frequency ratio $\omega_2/\omega_1 = 0.92$. The AD region is ascertained within the vertical lines. The width of the AD region along $\tau$-axis is larger using OCS than that using CCS.}
	\label{fig:rijke_fig54}
\end{figure}
In Fig.~\ref{fig:rijke_diff_kd}, we have employed OCS to the coupled Rijke tubes model using frequency ratio ($\omega_2/\omega_1$) as a control parameter at a fixed value of time delay $\tau = 0.5$~\cite{thomas18}. Unlike the previous two examples, AD region exists in two edges of the $\omega_2/\omega_1$-axis, and LCO is ascertained within the vertical lines. The AD regions are observed to enhance and come closer to each other using OCS than that using CCS (Fig.~\ref{fig:rijke_diff_kd}). Two different values of time-delay coupling strength parameter ($\alpha_{\tau}$) are chosen in the two subplots, and we have obtained similar results in both the cases. Besides, Fig.~\ref{fig:rijke_fig54} depicts the effect of time delay ($\tau$) on enhancing the AD region in the presence of a constant diffusive coupling strength (i.e., $\alpha_d =$ fixed) and a frequency ratio ($\omega_2/\omega_1 = 0.92$). Note that AD region is ascertained within the vertical lines. The two subplots correspond to $\alpha_d =0.12$ and $0.16$. The width of the AD regions enhances further along the $\tau$-axis for both values of $\alpha_d$ on using OCS. 
In short, the AD region enhances along the parameter axis of frequency ratio and time delay in the model of couple Rijke tubes. It is worth reiterating that a state of AD is preferred in a thermoacoustic system to suppress the oscillatory instability since oscillatory instabilities can be catastrophically detrimental to the performance and structural integrity of thermoacoustic systems such as those in rockets and gas turbine engines~\cite{sujith21}. With the implementation of occasional coupling, a wider range of parameters can be made available to bring about amplitude death in the system. 
Thus, we have studied the effect of OCS in coupled oscillators models. To be more explicit, we have employed the on-off coupling (i.e., through square wave function), an example of OCS, and ascertained that the AD regions enhances along the control parameter axis in all three models. Next, we choose a different form of $\chi(t)$ (other than that shown in Eq.~\ref{eq:ocs} or Fig.~\ref{fig:ocs_schematic}) and study its effect on AD. 

 \section{A different functional form of OCS: Half-wave rectified sinusoidal wave}
 \label{sec:other}  
 \begin{figure}[htbp!]
 	\hspace*{1 mm}
 	\includegraphics[width=50cm,height=4.5cm, keepaspectratio]{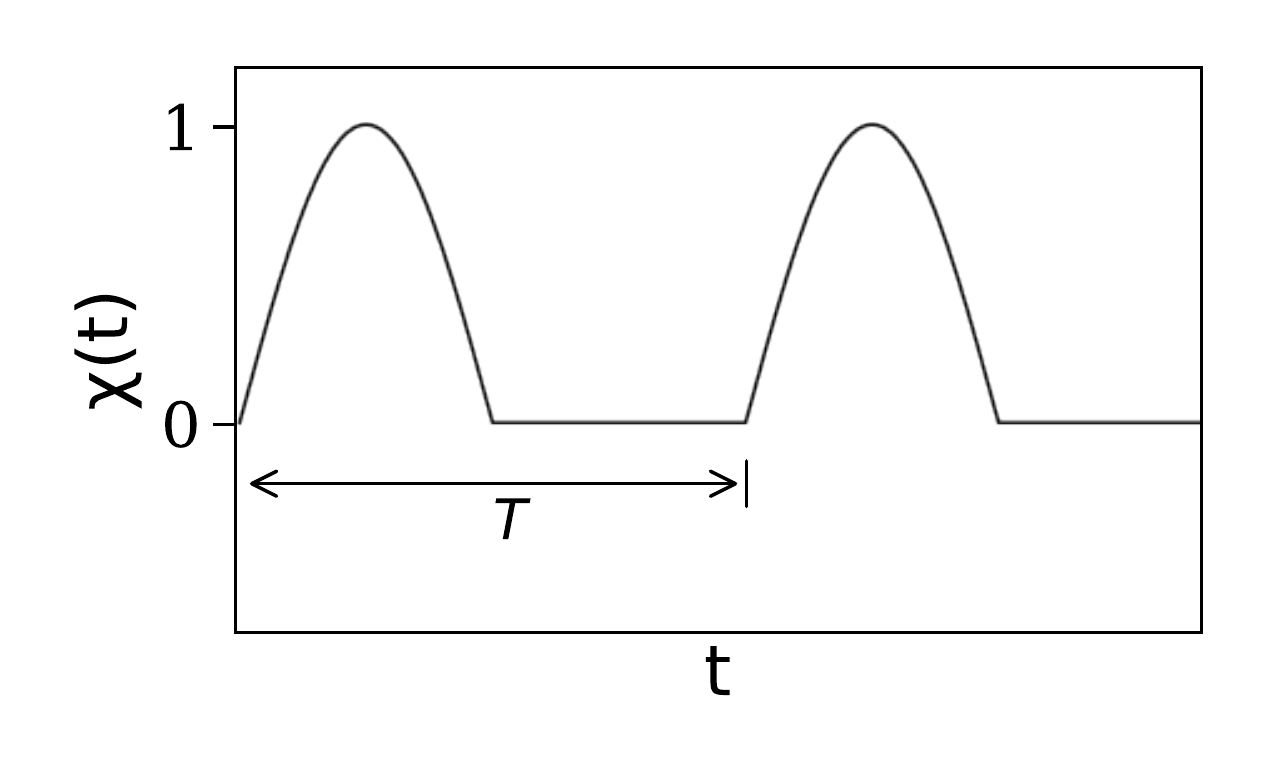}
 	\caption{The schematic diagram depicts the variation of the occasional coupling function $\chi(t)$ with time $t$ using the half-wave rectified sinusoidal wave (Eq.~\ref{eq:ocs_sinu}).} 
 	\label{fig:sinu_schematic}
 \end{figure}
 Implementing a square wave function in a mechanical system may not always be feasible as some finite time will always be required for the transition from on to off state for any coupling device (e.g., opening or closing a valve). Therefore, considering the gradual opening and closing of the valve, we adopt a half-wave rectified sinusoidal wave as the required functional form of $\chi(t)$, and mathematically, we can redefine $\chi(t)$ as follows:
 \begin{equation}
 \label{eq:ocs_sinu}
 \chi (t) := \sin^{+}(t) = 
 \begin{cases}
 \sin(\omega t),  \,  \, 2n\pi \leq t < (2n+1)\pi,\\
 0,  \, \, (2n-1)\pi \leq t < 2n\pi,
 \end{cases}
 \end{equation}
 where $\omega = 2\pi/T$ is the angular frequency of the sinusoidal wave and $T$ is the corresponding time period. Figure~\ref{fig:sinu_schematic} is the schematic diagram that shows the variation of $\chi(t)$ with time $t$ using Eq.~\ref{eq:ocs_sinu}.
 \begin{figure*}[htbp!]
 	\hspace*{0 cm}
 	\includegraphics[width=115cm,height=15.8cm, keepaspectratio]{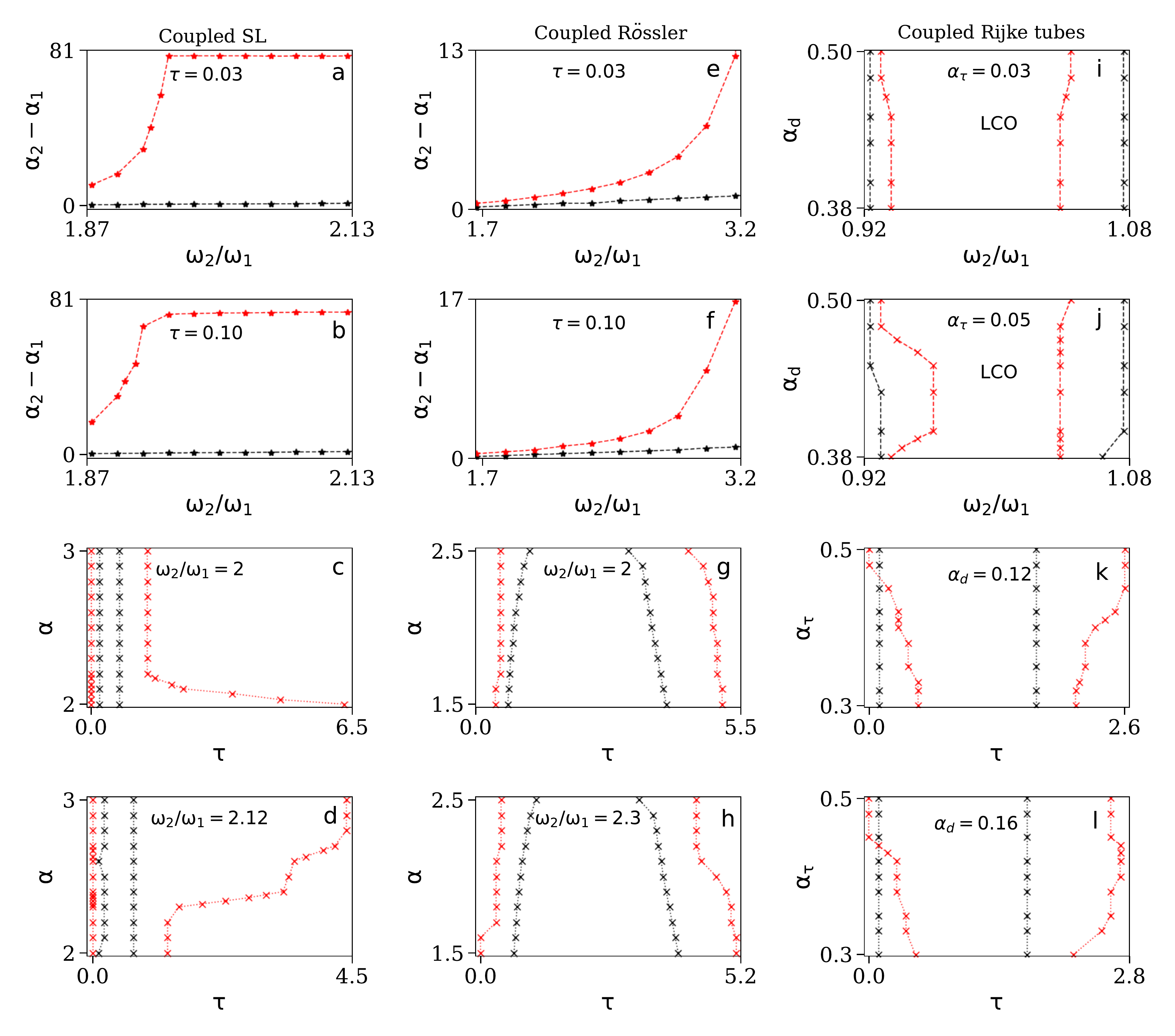}
 	\caption{\textit{(Color online)} The effect of OCS (Eq.~\ref{eq:ocs_sinu}) on AD is studied in three coupled oscillator models: SL, R\"ossler, and Rijke tube. The time-delay ($\tau$) and frequency ratio $(\omega_2 / \omega_1)$ are chosen as the control parameters. In all cases, the width of the AD regions are larger along the control parameter axes using OCS than that using CCS.}
 	\label{fig:sinu}
 \end{figure*}
 Figure~\ref{fig:sinu} is depicting the effect of OCS using Eq.~\ref{eq:ocs_sinu} on AD in all three examples of coupled oscillators discussed in this paper. The first, second, and third columns are corresponding to the coupled SL oscillators, coupled R\"ossler oscillators, and coupled Rijke tubes, respectively. Top two rows of each column use the frequency ratio ($\omega_2/ \omega_1$) as the control parameter and the bottom two rows use the time delay ($\tau$) as the required control parameter. For coupled SL oscillators and coupled R\"ossler oscillators, we have chosen two different $\tau$ in the top two rows and two different $\omega_2/\omega_1$ in the bottom two rows. In all cases, we obtain the favourable results, i.e., the employment of OCS enhances the AD regions along the control parameter axis. In both the examples, we choose the time period, $T = 1$ in Eq.~\ref{eq:ocs_sinu}.
 Furthermore, for coupled horizontal Rijke tubes, the top two rows correspond to $\tau = 0.5$ and two different values of the time delay coupling strength parameters ($\alpha_{\tau}$). We recall that the AD regions are at the two edges of Figs.~\ref{fig:sinu}i and \ref{fig:sinu}j. These AD regions come closer to each other in the presence of OCS. The bottom two rows (i.e., Figs.~\ref{fig:sinu}k and \ref{fig:sinu}l) correspond to the fixed frequency ratio $\omega_2 / \omega_1 = 0.92$ and two different values of the diffusive coupling strength parameters ($\alpha_d$). The width of the AD regions is observed to enhance with the increase in $\alpha_{\tau}$ along the $\tau$-axis. Also, at higher $\alpha_{\tau}$, the AD regions enhance on both sides of the $\tau$-axis. We choose the time period, $T = 2$ in Eq.~\ref{eq:ocs_sinu} for coupled Rijke tubes model.

 \section{Conclusion and Discussions}
 \label{sec:conclusion}   
 With an objective of enhancing the extent of the amplitude death regions along the control parameter axis, we have employed the OCS in time delay coupled oscillators. Towards that, first, the on-off coupling (i.e., through square wave function) has been employed. Our analysis has involved three examples of coupled oscillators: coupled SL oscillators, coupled R\"ossler oscillators, and coupled horizontal Rijke tubes. The horizontal Rijke tube is a prototypical model of a thermoacoustic system used to study the onset of thermoacoustic instability. Initially, the coupling strength parameter is chosen as the control parameter. It is observed that the AD regions enhance along the coupling strength parameter axis after employing the OCS compared to that using the CCS. \ag{We have performed a linear stability analysis for coupled SL oscillators to understand this enhancement analytically.} Next, motivated by the practical reality, we have adopted frequency ratio and time delay as the control parameters. Intriguingly, we have obtained favorable results, i.e., the enhancement of the width of the AD region, using the aforesaid control parameters. Finally, we have repeated our study with a different functional form of the OCS (half-wave rectified sinusoidal wave function) and got similar unaltered results. In short, this paper shows that the width of the amplitude death regions increases along the control parameter axis using the OCS. This finding can be helpful for a wide variety of physical systems such as thermoacoustic and aeroelastic systems, to name a few, where the presence of oscillations are hazardous.
 \ag{The linear stability analysis that we have performed for coupled SL oscillators to understand the enhancement of the AD region in the presence of the OCS analytically can not be stretched for coupled chaotic oscillators as chaotic dynamics are more complex than limit cycle dynamics. However, this analysis can be extended for the prototypical model of coupled Rijke tubes.} On the other hand, all the results have shown with a fixed combination of the on-off period ($T$) and on-off rate ($\theta$). Also, for simplicity, we choose $\theta = 0.5$ so that the time intervals over which the coupling is active and inactive become equal. The reported results may vary for different values of $T$ and $\theta$. Besides, from the perspective of experiments, the condition $T \gg T_s$ is preferable. For a larger value of $T$, however, the occasional coupling may not always be a recommended tool to enhance the amplitude death regions along the control parameter axis. As a possible future direction, a detailed study on the effectiveness of the on-off coupling scheme in time-delay coupled oscillators using different combinations of $(T, \theta)$ is an exciting direction to pursue.
 Furthermore, we may extend our study to the slow-fast Fitzhugh-Nagumo~\cite{fitzhugh61} or the Hodgkin-Huxley~\cite{hodgkin52} oscillators, where the dynamics change a lot on short timescales (spikes). The employment of the on-off coupling in these oscillators in the context of amplitude death and studying the consequences of different combinations of $(T, \theta)$ are interesting. In such cases, one might need to change $\theta$ or achieve different results, eventually ending up in different attractors. Finally, there are several other examples of the occasional coupling schemes available in the literature, and a brief review of such schemes has been done by Ghosh and Chakraborty~\cite{ghosh20}. Those schemes have been reported mostly in the context of synchronization; implementation of such occasional coupling schemes in the purview of amplitude death may yield interesting results.
\section*{Acknowledgments}
The authors thank Tutun Hazra for making the schematic diagram of coupled Rijke tubes. The authors also thank Ankan Banerjee, Pijush Pandey, Sneha Srikanth, and Somnath De for their fruitful comments. A special thank to Awadhesh Prasad from the University of Delhi, for his help during the revision of this paper. A. G. gratefully acknowledges the Institute Post-Doctoral Fellowship of Indian Institute of Technology Madras, India. R. I. S. expresses his gratitude to the Department of Science and Technology, Government of India, for providing financial support under Grant Number JCB/2018/000034/SSC (J. C. Bose Fellowship).
\section*{Data Availability Statements}
The data that support the findings of this study are available from the corresponding author upon reasonable request.

\section*{Conflict of interest}
The authors declare that they have no known competing financial interests or personal relationships that could have appeared to influence the work reported in this paper.
\appendix
\section{Mathematical model of the horizontal Rijke tube}
\label{sec:appen_rijke}
Here, we focus on the mathematical model of an uncoupled horizontal Rijke tube~\cite{balasubramanian08}. This model is developed from the linearized momentum and linearized energy equations of the acoustic field with the approximation of zero Mach number and neglecting the mean temperature gradient~\cite{balasubramanian08}. The non-dimensionalized form of the governing equations are:
\begin{eqnarray}\label{eq:single_rijke}
&& \gamma M \frac{\partial u}{\partial t} +  \frac{\partial p}{\partial x} =0, \label{eq:single_rijke_a}\\
&& \frac{\partial p}{\partial t} + \gamma M \frac{\partial u}{\partial x} + \zeta p = (\gamma-1)\cdot \dot{Q}(t) \cdot \delta(x - x_f),\label{eq:single_rijke_b}
\end{eqnarray}
where $p$ and $u$ are the pressure fluctuation and the velocity fluctuation respectively in the duct. The parameters: $\gamma$, $M$, and $\zeta$, are the ratio of the specific heats in the medium, the Mach number of the flow, and the damping coefficient, respectively. $\dot{Q}(t)$ is the source term which is located at a spatial distance $x_f$, and the dot on $Q$ represents the time derivative. More explicitly, $\dot{Q}(t)$ measures the heat release rate per unit area. Lastly, $\delta(\cdot)$ represents the standard Dirac delta function. The explicit form of $\dot{Q}(t)$ is given by 
\begin{eqnarray}\label{eq:source}
\dot{Q}(t) =&& \frac{2 L_w (T_w - \bar{T})}{\sqrt{3}Sc_0 \bar{p}} \sqrt{\pi \lambda C_v u_0 \bar{\rho}l_c} \nonumber \\ &&\times \left[ \sqrt{|{\frac{1}{3} + u_f(t-\tau_1)}|} - \sqrt{\frac{1}{3}}\right].  
\end{eqnarray}
In the above equation (Eq.~\ref{eq:source}), $L_w$, $T_w$, and $l_c$ are the length, temperature, and radius of the wire-mesh respectively; $S$, $c_0$, and $\bar{p}$ are the cross-sectional area of the tube, velocity of sound, and ambient pressure, respectively; $\lambda$, $C_v$, and $u_0$ are the thermal conductivity of the medium within the duct, specific heat at constant volume of the medium within the duct, and steady state velocity of the flow, respectively; $\bar{\rho}$ is the mean density of the medium within the tube. The last term, $u_f(t-\tau_1)$, physically implies that due to the thermal inertia of the medium, the heat release rate at the wire-mesh gets delayed by a constant time lag ($\tau_1$) at the boundary.   
We choose the boundary condition that the acoustic pressure at two ends of the duct are identical with the ambient pressure, i.e., $p(0, t) = p(1, t) = 0$. For simplicity, we may transform the partial differential equations (Eqs.~\ref{eq:single_rijke_a} and \ref{eq:single_rijke_b}) into ordinary differential equations using the Galerkin technique~\cite{lores73}. Following this Galerkin method, we may write $p$ and $u$ as follow:
\begin{eqnarray}\label{eq:transform}
&& u = \sum_{k = 1}^{N} \eta_k \cos(k\pi x), \\ &&p = -\sum_{k = 1}^{N} \dot{\eta}_k \frac{\gamma M}{k \pi}\sin(k\pi x).  
\end{eqnarray}
Here, $\eta_k$ and $\dot{\eta}_k$ represent the coefficients of the acoustic velocity ($u$) and acoustic pressure ($p$), respectively. $N$ is the total number of modes, and for this example $N = 10$ is sufficient to get a suitable solution~\cite{subramanian10}. Thus, the equations of motion of the Rijke tube in terms of the Galerkin modes are given by:
\begin{subequations}
	\label{eq:single_rijke_trans}
	\begin{eqnarray}
	\frac{d \eta_k}{d t} &=& \dot{\eta}_k, \label{eq:single_rijke_trans_a}\\
	\frac{d \dot{\eta}_k}{d t} + 2\zeta_k \omega_k \dot{\eta}_k + \omega^2_k {\eta}_k &=&  - k \pi K \left[ \sqrt{|{\frac{1}{3} + u_f(t-\tau_1)}|} - \sqrt{\frac{1}{3}}\right] \nonumber \\ && \times \sin(k \pi x_f) \label{eq:single_rijke_trans_b},  
	\end{eqnarray}
\end{subequations}
where 
\begin{equation}
u_f(t-\tau_1) = \sum_{k = 1}^{N} \eta_k (t-\tau_1) \cos(k\pi x),
\end{equation}
and $\omega_k = k \pi$, the angular frequency of the $k^{\rm th}$ mode. $2\zeta_k \omega_k \dot{\eta}_k$ is the damping term, and the parameter $\zeta_k$ is defined as: 
\begin{equation}
\zeta_k = \frac{1}{2 \pi} \left[c_1 \frac{\omega_k}{\omega_1} + c_2 \sqrt{\frac{\omega_1}{\omega_k}}\right].
\end{equation}
Parameters $c_1$ and $c_2$ are the damping coefficients. $K$ is the heater power. As we increase $K$ from zero, the Rijke tube goes through a subcritical Hopf bifurcation at $K_{\rm Hopf} = 0.62$~\cite{thomas18}, i.e., the stable fixed point loses its stability and forms a limit cycle. Thus, in this study, we choose a value of $K$ which satisfy the condition $K > K_{\rm Hopf}$. Thus, we obtain the equations of motion of the uncoupled Rijke tube (Eq.~\ref{eq:single_rijke_trans}). In order to solve Eq.~\ref{eq:single_rijke_trans} numerically, we have chosen the initial conditions as $\eta_1(0) = 0.01$ and $\dot{\eta}_1(0) = 0.001$; rest nine modes of $\eta_k (t)$ and $\dot{\eta}_k(t)$ have been adopted as zero initially~\cite{thomas18}. The values of the parameters chosen in this study are enlisted in Table~\ref{table:1}.
\begin{table}[htbp!]
	\caption{\label{table:1} The parameter values enlisted in this table are chosen for the simulation of coupled Rijke tubes~\cite{thomas18}.}
	\begin{center}
		\begin{tabular}{ lccccccccccccr}
			\hline \hline
			Parameter &&&&&&&& Corresponding value \\
			\hline 
			$M$ &&&&&&&& $0.01$    \\
			$x_f$ &&&&&&&& $0.25$  \\
			$c_1$ &&&&&&&& $0.10$ \\
			$c_2$  &&&&&&&& $0.06$ \\
			$K$ &&&&&&&& $0.92$  \\
			$\tau_1$ &&&&&&&& $0.20$    \\
			$\gamma$ &&&&&&&& $1.40$ \\
			\hline \hline
		\end{tabular}
	\end{center}
\end{table}

\bibliography{Ghosh_etal_Manuscript.bib}
\end{document}